\documentclass[a4paper,runningheads]{llncs}
\usepackage{sosy-paper}
\usepackage[textsize=tiny]{todonotes}
\usepackage{fancyvrb}
\usepackage{fvextra}
\usepackage{longtable}
\usepackage{tcolorbox}
\usepackage{marvosym}

\definetool{\testcomp}{Test-Comp}
\definetool{\rers}{RERS}
\newcommand{\correspondinglink}{mailto:dirk.beyer@sosy.ifi.lmu.de}
\newcommand{\onlinetutoriallink}{\url{https://cpachecker.sosy-lab.org/doc.php}\xspace}
\newcommand{\corresponding}{\href{\correspondinglink}{\raisebox{4.7pt}{\Letter}}}

\usepackage[paperwidth=155mm,
            paperheight=235mm,
            left=16mm,
            right=16mm,
            top=17mm,
            bottom=25mm
           ]{geometry}

\title{
  Software Verification with \let\toolnamesize\relax\cpachecker 3.0:
  \\ Tutorial and User Guide
  \\ (Extended Version)
}
\titlerunning{Software Verification with \let\toolnamesize\relax\cpachecker 3.0: Tutorial and User Guide}
\author{
    Daniel~Baier\orcidID{0000-0001-9116-1974},
    Dirk~Beyer\orcidID{0000-0003-4832-7662}\corresponding{},
    Po-Chun~Chien\orcidID{0000-0001-5139-5178},
    Marie-Christine~Jakobs\orcidID{0000-0002-5890-4673},
    Marek~Jankola\orcidID{0009-0008-7961-190X},
    Matthias~Kettl\orcidID{0000-0001-7365-5030},
    Nian-Ze~Lee\orcidID{0000-0002-8096-5595},
    Thomas~Lemberger\orcidID{0000-0003-0291-815X},
    Marian~Lingsch-Rosenfeld\orcidID{0000-0002-8172-3184},
    Henrik~Wachowitz\orcidID{0000-0002-4768-4054}, and
    Philipp~Wendler\orcidID{0000-0002-5139-341X}
}
\authorrunning{\mbox{\tiny
Baier,
Beyer,
Chien,
Jakobs,
Jankola,
Kettl,
Lee,
Lemberger,
Lingsch-Rosenfeld,
Wachowitz,
Wendler}}
\institute{LMU Munich, Munich, Germany\\[5mm]
\scalebox{.12}{
\begin{tikzpicture}[y=0.80pt, x=0.80pt, yscale=-1.000000, xscale=1.000000, inner sep=0pt, outer sep=0pt]
\begin{scope}[shift={(-221.97172,-411.62445)}]
  \begin{scope}[shift={(174.94262,35.80334)}]
    \path[fill=cpacheckergreen] (343.2732,431.6103) .. controls (346.6213,431.6104) and
      (349.1539,434.3576) .. (350.8709,439.8519) .. controls (354.3049,450.1540) and
      (356.7516,455.3050) .. (358.2111,455.3049) .. controls (359.3271,455.3050) and
      (360.0397,454.0001) .. (361.2416,452.2830) .. controls (385.3654,413.6505) and
      (449.4021,378.7382) .. (451.7061,377.1931) .. controls (453.5563,375.7472) and
      (455.2449,375.7948) .. (457.7931,375.8348) .. controls (460.5402,375.8350) and
      (465.3567,378.5405) .. (466.3841,379.8938) .. controls (467.9207,382.1169) and
      (468.9861,381.3650) .. (470.1138,389.0872) .. controls (468.5352,393.5092) and
      (467.3858,393.7940) .. (466.3207,394.8515) .. controls (427.1214,422.0592) and
      (392.8516,441.3973) .. (364.5211,485.6959) .. controls (362.5465,488.7865) and
      (358.5115,490.3318) .. (352.4162,490.3318) .. controls (346.2350,490.3318) and
      (342.5863,490.0743) .. (341.4703,489.5591) .. controls (338.5514,488.2714) and
      (335.1174,481.7039) .. (331.1683,469.8565) .. controls (326.7041,456.7215) and
      (324.4719,448.4799) .. (324.4720,445.1317) .. controls (324.4719,441.5261) and
      (327.4767,438.0491) .. (333.4863,434.7009) .. controls (337.1778,432.6406) and
      (340.4401,431.6104) .. (343.2732,431.6103);
    \path[fill=sosyblue] (119.3807,446.3865) -- (147.2947,454.8240) .. controls
      (145.4196,462.6522) and (142.4665,469.1912) .. (138.4354,474.4412) .. controls
      (134.4040,479.6912) and (129.3884,483.6521) .. (123.3885,486.3240) .. controls
      (117.4353,488.9959) and (109.8415,490.3318) .. (100.6072,490.3318) .. controls
      (89.4041,490.3318) and (80.2400,488.7146) .. (73.1150,485.4803) .. controls
      (66.0369,482.1990) and (59.9197,476.4568) .. (54.7635,468.2537) .. controls
      (49.6072,460.0506) and (47.0291,449.5506) .. (47.0291,436.7537) .. controls
      (47.0291,419.6913) and (51.5525,406.5897) .. (60.5994,397.4490) .. controls
      (69.6931,388.2616) and (82.5369,383.6679) .. (99.1307,383.6678) .. controls
      (112.1150,383.6679) and (122.3103,386.2929) .. (129.7166,391.5428) .. controls
      (137.1696,396.7928) and (142.7009,404.8553) .. (146.3104,415.7303) --
      (118.1854,421.9881) .. controls (117.2009,418.8475) and (116.1697,416.5506) ..
      (115.0916,415.0974) .. controls (113.3103,412.6600) and (111.1306,410.7850) ..
      (108.5525,409.4724) .. controls (105.9744,408.1600) and (103.0915,407.5038) ..
      (99.9041,407.5037) .. controls (92.6853,407.5038) and (87.1541,410.4100) ..
      (83.3104,416.2224) .. controls (80.4041,420.5350) and (78.9509,427.3084) ..
      (78.9510,436.5428) .. controls (78.9509,447.9803) and (80.6853,455.8318) ..
      (84.1541,460.0974) .. controls (87.6228,464.3162) and (92.4978,466.4256) ..
      (98.7791,466.4256) .. controls (104.8728,466.4256) and (109.4665,464.7147) ..
      (112.5604,461.2928) .. controls (115.7009,457.8709) and (117.9743,452.9022) ..
      (119.3807,446.3865);
    \path[fill=uniorange] (152.5188,387.2537) -- (205.4641,387.2537) .. controls
      (216.9953,387.2538) and (225.6203,389.9960) .. (231.3391,395.4803) .. controls
      (237.1046,400.9647) and (239.9874,408.7694) .. (239.9875,418.8943) .. controls
      (239.9874,429.3006) and (236.8468,437.4334) .. (230.5657,443.2928) .. controls
      (224.3312,449.1522) and (214.7922,452.0819) .. (201.9485,452.0818) --
      (184.5110,452.0818) -- (184.5110,490.3318) -- (152.5188,490.3318) --
      (152.5188,387.2537)(184.5110,431.1990) -- (192.3157,431.1990) .. controls
      (198.4562,431.1991) and (202.7687,430.1444) .. (205.2532,428.0349) .. controls
      (207.7375,425.8788) and (208.9797,423.1366) .. (208.9797,419.8084) .. controls
      (208.9797,416.5741) and (207.9016,413.8319) .. (205.7454,411.5818) .. controls
      (203.5891,409.3319) and (199.5344,408.2069) .. (193.5813,408.2068) --
      (184.5110,408.2068) -- (184.5110,431.1990);
    \path[fill=unigrey] (294.4918,473.3162) -- (258.2105,473.3162) --
      (253.2183,490.3318) -- (220.6636,490.3318) -- (259.4058,387.2537) --
      (294.1402,387.2537) -- (332.8824,490.3318) -- (299.5543,490.3318) --
      (294.4918,473.3162)(287.8121,451.0271) -- (276.4214,413.9724) --
      (265.1011,451.0271) -- (287.8121,451.0271);
  \end{scope}
\end{scope}

\end{tikzpicture}
} \quad
\url{https://cpachecker.sosy-lab.org}
\vspace{-5.5mm}
}

\definecolor{thomas}{HTML}{ba84bf}

\definecolor{matthias}{HTML}{7a4300}

\definecolor{marek}{HTML}{ff7e00}

\definecolor{marian}{HTML}{42a5f5}

\definecolor{marie}{HTML}{00ffff}

\newcommand{\mypaperkeywords}{
    CPAchecker \and
    Configurable Program Analysis \and
    Formal Verification \and
    Model Checking \and
    Software Verification \and
    Program Analysis \and
    Testing \and
    Tutorial \and
    Correctness Certification \and
    Witnesses \and
    Witness Validation \and
    Fault Visualization
}
\DisableLigatures{family=tt*}
\newcommand{\cpaconfig}[1]{\href{https://svn.sosy-lab.org/software/cpachecker/tags/cpachecker-3.0/config/#1.properties}{\texttt{-{}-#1}}}
\newcommand{\cpaspec}[1]{\href{https://svn.sosy-lab.org/software/cpachecker/tags/cpachecker-3.0/config/specification/#1.spc}{\texttt{#1}}}
\newcommand{\exampleprog}[1]{\href{https://svn.sosy-lab.org/software/cpachecker/tags/cpachecker-3.0/doc/examples/#1}{\texttt{#1}}}
\definetool{\gzip}{Gzip}

\lstdefinestyle{spc}{
    language=C,
    basicstyle=\lstbasicfontsize\ttfamilywithbold,
    keywordstyle=\color{blue},
    commentstyle=\color{green!50!black},
    stringstyle=\color{red},
    breaklines=true,
    showstringspaces=false,
    tabsize=2,
    morekeywords={OBSERVER, AUTOMATON, INITIAL, STATE, USEFIRST, MATCH, STOP, END}
}
\renewcommand\lstbasicfontsize\scriptsize

\newcommand{\formula}[2]{\tikz[baseline]{\node[shape=rectangle,line width=1pt,draw=#2,fill=#2!30,inner sep=1pt] at (0,.64ex){\hspace{.2em}\texttt{\strut#1}\hspace{.1em}\strut};}}
\newcommand{\examplecodesize}{\fontsize{7}{8.5}\selectfont}

\definecolor{tolDarkBlue}{HTML}{332288}
\definecolor{tolDarkGreen}{HTML}{117733}
\definecolor{tolTeal}{HTML}{44aa99}
\definecolor{tolLightBlue}{HTML}{88ccee}
\definecolor{tolYellow}{HTML}{ddcc77}
\definecolor{tolRed}{HTML}{cc6677}
\definecolor{tolViolet}{HTML}{aa4499}
\definecolor{tolBurgundy}{HTML}{882255}
\definecolor{tolPastelTeal}{HTML}{d0eae6}
\definecolor{tolPastelGreen}{HTML}{c4ddcc}
\definecolor{tolPastelBlue}{HTML}{e1f2fb}
\definecolor{tolPastelRed}{HTML}{f2d9dd}
\definecolor{tolPastelYellow}{HTML}{f6f2dd}
\definecolor{tolPastelViolet}{HTML}{ead0e6}
\definecolor{tolPastelBurgundy}{HTML}{e1c8d5}
\definecolor{colorAlmostWhite}{HTML}{f5f7f7}

\tikzset{
    st/.style={
        font=\ttfamily,
        shape=rectangle,
        rounded corners=.5em,
        fill=gray!40,
        inner xsep=.3em,
        inner ysep=0em,
        text height=2ex,
        text depth=.6ex
    },
    actor edge/.style={
        draw=black,
        ->,
        shorten >=2pt,
        shorten <=2pt,
        >=latex, 
    },
    obj/.style={
        draw,
        align=center,
        minimum height=1.1cm,
        font=\sffamily,
    },
    actor/.style={
        obj,
        fill=colorAlmostWhite,
    },
    artifact/.style={
        obj,
        draw=black,  %
        minimum height=.5cm,
        minimum width=.5cm,
        inner sep=2pt,
        fill=colorAlmostWhite,
        document,
        rounded corners=1pt,
    },
    artifact label/.style={
        font=\sffamily\scriptsize,
        inner sep=0pt,
        color=black,
    },
    program/.style={
        artifact,
        fill=tolPastelBlue,
    },
    spec/.style={
        artifact,
        fill=tolPastelYellow,
    },
    config/.style={
        artifact,
        fill=tolPastelTeal,
    },
    verdict/.style={
        artifact,
        left color=tolPastelGreen,
        right color=tolPastelRed,
    },
    report/.style={
        artifact,
        shading=axis, shading angle=90,
        left color=gray,
        right color=white,
    },
    witness/.style={
        artifact,
        fill=tolPastelBurgundy,
    },
    testcase/.style={
        artifact,
        fill=tolPastelRed,
    }
}
\definecolor{itpGreen}{rgb}{0,1,0}

\lstdefinestyle{yaml}{
    basicstyle=\color{blue}\lstbasicfontsize\ttfamilywithbold,
    rulecolor=\color{black},
    comment=[l]{:},
    commentstyle=\color{black},
    string=[s]{"}{"},
    stringstyle=\color{green},
    numberstyle=\color{black}
}

\definecolor{startButtonColor}{HTML}{ffc107}
\definecolor{prevNextButtonColor}{HTML}{28a745}
\definecolor{buttonBorderColor}{HTML}{222222}
\definecolor{vButtonColor}{HTML}{cccccc}
\definecolor{tabButtonColor}{HTML}{007bff}
\newcommand{\inlinebutton}[3][black]{\tikz[baseline]{\node[anchor=base, rounded corners=1pt,fill=#3, inner sep=3pt,font=\sffamily\small]{\textcolor{#1}{#2}};}}
\newcommand{\reporttab}[1]{\inlinebutton[white]{#1}{tabButtonColor}}

\newcommand{\exonline}[1]{(cf.~example \mbox{\href{https://vcloud.sosy-lab.org/cpachecker/webclient/run/example/#1}{#1}} on the web service)\xspace}

\newcommand{\onWebclient}[2]{\href{https://vcloud.sosy-lab.org/cpachecker/webclient/run/example/#1}{#2}\xspace}

\definetool{\fmtools}{FM-Tools}
\definetool{\fmweck}{FM-Weck}
\definetool{\coveriteam}{CoVeriTeam}
\definetool{\coveriteamservice}{CoVeriTeam Service}

\RequirePackage{graphicx}
\usepackage[firstpageonly=true,angle=0,vpos=.005\paperheight,hpos=.01\linewidth,vanchor=t,hanchor=l]{draftwatermark}
\SetWatermarkText{%
\raisebox{0cm}
{\hspace{0.92\linewidth}
\href{https://doi.org/10.5281/zenodo.13612338}{\includegraphics[width=11mm]{badges/FM\_2024\_AE\_reusable}}
\href{https://doi.org/10.5281/zenodo.13612338}{\includegraphics[width=11mm]{badges/FM\_2024\_AE\_available}}
}}

\begin{document}
\acrodef{CEGAR}[CEGAR]{counterexample-guided abstraction refinement}
\maketitle
\blfootnote{This is an extended version of the user guide published at FM~\cite{CPAchecker-3.0}.}

\begin{abstract}
  This tutorial provides an introduction to \cpachecker for users.
\cpachecker is a flexible and configurable framework for software verification and testing.
The framework provides many abstract domains,
such as
BDDs,
explicit values,
intervals,
memory graphs, and
predicates,
and many program-analysis and model-checking algorithms,
such as
abstract interpretation,
bounded model checking,
\impact,
interpolation-based model checking,
\kinduction,
PDR,
predicate abstraction, and
symbolic execution.
This tutorial presents basic use cases for \cpachecker in formal software verification, focusing
on its main verification techniques with their strengths and weaknesses.
It also shows further use cases of \cpachecker for test-case generation
and witness-based result validation.
The envisioned readers are assumed to possess a background in automatic formal verification and program analysis,
but prior knowledge of \cpachecker is not required.
This tutorial and user guide is based on \cpachecker in version~3.0.
This user guide's latest version and other documentation are available at \onlinetutoriallink.

\end{abstract}

\section{Introduction}
\label{sec:intro}

\begin{figure}[t]
    \centering
	\begin{tikzpicture}[node distance=5mm]

    \node[] (program) {
        \hyperref[sec:programs]{\tikz{
            \node[program] (program-obj) {};
            \node[artifact label, below = 2pt of program-obj.south] (program-label) {Program};
        }}
    };

    \node[below right = 1mm and -1mm of program.east] (specification) {
        \hyperref[sec:specification]{\tikz{
            \node[spec] (spec-obj) {};
            \node[artifact label, below = 2pt of spec-obj.south] (spec-label) {Specification};
        }}
    };

    \node[below = 1mm of program] (configuration) {
        \hyperref[sec:config]{\tikz{
            \node[config] (conf-obj) {};
            \node[artifact label, below = 2pt of conf-obj.south] (conf-label) {Configuration};
        }}
    };
    \node[below = 1mm of specification] (witness-in) {
        \hyperref[sec:output-witness]{\tikz{
            \node[witness] (witness-obj) {};
            \node[artifact label, below = 2pt of witness-obj.south] (witness-label) {Witness};
        }}
    };

    \coordinate (left-middle) at ($(program)!.5!(witness-in)$);

    \node[actor, draw=none, fill=white, right = 10mm of left-middle -| specification.east] (CPAchecker) {\scalebox{.25}{\input{logos/cpa.tex}}};

    \coordinate[right = 20mm of CPAchecker] (right-middle);
    \coordinate (start-output-artifacts) at (right-middle |- program.north);

    \node[anchor=north west] (verdict) at (start-output-artifacts) {
        \hyperref[sec:verdict]{\tikz{
            \node[inner sep=0pt] (verdict-obj) {\rotatebox{-90}{\scalebox{.5}{\begin{tikzpicture}
                \draw[black, fill=gray!50, rounded corners=1pt] (-0.35, 0.35cm) rectangle (1.35, -0.35);
                \draw[fill=red] (0,0) circle (0.20cm);
                \draw[fill=yellow] (0.50,0) circle (0.20cm);
                \draw[fill=green] (1.0,0) circle (0.20cm);
            \end{tikzpicture}}}};
            \node[artifact label, below = 2pt of verdict-obj.south] (verdict-label) {Verdict};
        }}
    };

    \node[below left = 2mm and -1mm of verdict.west] (report) {
        \hyperref[sec:report]{\tikz{
            \node[report] (report-obj) {};
            \node[artifact label, below = 2pt of report-obj.south] (report-label) {Report};
        }}
    };
    \node[below = 1mm of verdict] (witness-out) {
        \hyperref[sec:output-witness]{\tikz{
            \node[witness] (witness-obj) {};
            \node[artifact label, below = 2pt of witness-obj.south] (witness-label) {Witness};
        }}
    };
    \node[below = 1mm of report] (testcase) {
        \hyperref[sec:output-tests]{\tikz{
            \node[testcase] (testcase-obj) {};
            \node[artifact label, below = 2pt of testcase-obj.south] (testcase-label) {Tests};
        }}
    };

    \draw[actor edge] (program.east) -- ($(CPAchecker.west) + (0, 10pt)$);
    \draw[actor edge] (specification.east) -- ($(CPAchecker.west) + (0, 2.5pt)$);
    \draw[actor edge] (configuration.east) -- ($(CPAchecker.west) + (0, -2.5pt)$);
    \draw[actor edge] (witness-in.east) -- ($(CPAchecker.west) + (0, -10pt)$);
    \draw[actor edge] ($(CPAchecker.east) + (0, 10pt)$) -- (verdict.west);
    \draw[actor edge] ($(CPAchecker.east) + (0, 2.5pt)$) -- (report.west);
    \draw[actor edge] ($(CPAchecker.east) + (0, -2.5pt)$) -- (witness-out.west);
    \draw[actor edge] ($(CPAchecker.east) + (0, -10pt)$) -- (testcase.west);
\end{tikzpicture}
    \vspace*{-4mm}
    \caption{
        Inputs and outputs of \cpachecker
        when it is used as a \hyperref[sec:algorithms]{verifier},
        witness validator,
        or test-case generator
        }
    \label{fig:inputsOutputs}
    \vspace*{-5mm}
\end{figure}

\cpachecker~\cite{CPACHECKER} is a framework
for configurable software verification with a focus on the verification of C programs.
It is based on the concept of configurable program analysis~\cite{CPA,CPAplus,HBMC-dataflow}
and provides an extensive collection of verification algorithms and abstract domains.
Throughout the past years,
\cpachecker has been a top contender in the International Competition on Software Verification~\cite{SVCOMP22,SVCOMP23,SVCOMP24}
and has helped identify \href{https://svn.sosy-lab.org/software/cpachecker/tags/cpachecker-3.0/doc/Achievements.md#bugs-found-with-cpachecker}
{over 240~bugs in Linux device drivers}~\cite{LDV-Toolset,LDV,LDV12}.

\cpachecker is open source and written in Java.
Founded in~%
\href{https://svn.sosy-lab.org/trac/cpachecker/changeset/19155/CPAchecker}{2007}
at Simon Fraser University, it is now maintained by an active community
(project statistics can be found on \href{https://openhub.net/p/cpachecker}{OpenHub.net}).
It puts a high priority on extensibility and flexible reuse of components for developers.
The architecture and features of the framework are described in other articles~\cite{CPACHECKER,CPAchecker-ASV}.
More information about the achievements, history, and license of \cpachecker are available in
\cref{appendix:projectInfo}.

\subsection{Use Cases of \cpachecker}

There are three main use cases of \cpachecker,
with their inputs and outputs summarized in \cref{fig:inputsOutputs}:
(1)~As a \emph{verifier}, \cpachecker takes as input a program and a specification,
and returns a verdict, a verification report, and a verification witness.
The verdict specifies whether the given program adheres to the specification,
the verification report allows users to examine the verification result, and
the witness contains a machine-readable justification for the returned verdict.
(2)~As a \emph{witness validator}~\cite{WitnessesJournal,VerificationWitnesses-2.0},
\cpachecker takes as input a program, a specification, and a witness,
and returns a verdict
that indicates whether the witness could be confirmed by \cpachecker.
(3)~As a \emph{test-case generator}~\cite{CoVeriTest-STTT,MultiGoal-ProductLines,HYBRIDTIGER-TESTCOMP20},
\cpachecker takes as input a program and a test-coverage specification,
and returns a set of test cases that cover the program according to the specification.

\cpachecker is also used for program transformation~\cite{ReducerCMC,DiffCond,FRED,LoopAbstractionCEGAR,CEGAR-PT},
to explore decompositions of verification problems~\cite{CCEGAR,DecompositionOfSpecifications,DSS-PACMSE},
and to parallelize verification approaches~\cite{BAM-parallel,DSS-PACMSE}.
This tutorial focuses on using \cpachecker as a verifier.
Information about \cpachecker as a witness validator and test-case generator
is presented in \cref{appendix:witness-validation} and \cref{appendix:test-gen} respectively.

\subsection{Configurable Program Analysis}
\cpachecker uses configurable program analysis (CPA)~\cite{CPA,CPAplus,HBMC-dataflow} to
compute a program's reachable states.
A CPA specifies an abstract domain and
a precision used to explore a program's reachable states.
The abstract domain defines the representation of a program's state,
while the precision defines how precise the abstraction should be.
Various CPAs have been implemented in \cpachecker,
each tailored to handle specific program features and perform a dedicated analysis.
CPAs can also be combined to achieve synergy.
Furthermore, precisions can be adjusted dynamically~\cite{CPAplus},
making an analysis coarse but efficient, or
precise but resource-consuming.
\cpachecker automatically adjusts the precisions via
\cegar~\cite{ClarkeCEGAR,CPAexplicit,RefinementSelection,BAM-COW-Refinement}
or some carefully-designed procedures~\cite{CPA-DF}.

\subsection{Documentation and Communication}

The \href{https://svn.sosy-lab.org/software/cpachecker/tags/cpachecker-3.0/README.md}{\texttt{README}}
and directory
\href{https://svn.sosy-lab.org/software/cpachecker/tags/cpachecker-3.0/doc}{\texttt{doc/}}
in the \cpachecker project provide useful information for users and developers.
For an overview on the architecture,
we recommend the tool paper~\cite{CPACHECKER} on \cpachecker
and the publications regarding the CPA concept~\cite{CPA,CPAplus,HBMC-dataflow}.
\cpachecker supports various verification algorithms and techniques.
The most important techniques in \cpachecker are explained
in separate publications, including
data-flow and value analysis~\cite{
    HBMC-dataflow, %
    CPAexplicit, %
    CPA-DF %
},
SMT-based verification algorithms~\cite{
    AlgorithmComparison-JAR, %
    CPAsymexec, %
    IMC-JAR %
},
block-abstraction memoization~\cite{
    BAM,
    BAMInterprocedural, %
    BAM-parallel, %
    BAM-COW-Refinement %
},
program transformations~\cite{ReducerCMC,DiffCond,FRED,LoopAbstractionCEGAR,CEGAR-PT},
cooperative verification~\cite{kInduction,IMCDF},
witness certification and validation~\cite{
    WitnessesJournal, %
    VerificationWitnesses-2.0 %
}, and
test-case generation~\cite{CoVeriTest-STTT,MultiGoal-ProductLines,HYBRIDTIGER-TESTCOMP20}.
The configurations of \cpachecker that were submitted to competitions
are described in the competition contribution papers of
\svcomp~\cite{
    CPACHECKERABE-SVCOMP12,
    CPACHECKERMEMO-SVCOMP12,
    CPACHECKEREXPLICIT-SVCOMP13,
    CPACHECKERSEQCOM-SVCOMP13,
    CPACHECKER-SVCOMP14,
    CPACHECKER-SVCOMP15,
    CPABAM-SVCOMP16,
    CPABAM-SVCOMP17,
    PESCO-SVCOMP19,
    CPALOCKATOR-SVCOMP21,
    GRAVES-SVCOMP22,
    CPACHECKER-SVCOMP24
}, \testcomp~\cite{
    COVERITEST-TESTCOMP21,
    COVERITEST-TESTCOMP20,
    COVERITEST-TESTCOMP19,
    COVERITEST
}, and \rers~\cite{
    CPABDD,
    CPABDD-memics
}.
These publications give an indication
of the breadth of analyses available in \cpachecker
and its power and flexibility as a verification framework.

Questions, bug reports, and feature requests for \cpachecker
are always welcome on its mailing list ({\smaller\url{https://groups.google.com/g/cpachecker-users}})
and the issue tracker ({\smaller\url{https://gitlab.com/sosy-lab/software/cpachecker/-/issues}}).

\subsection{CPAchecker in Education}
Due to the many algorithms and abstract domains, and the clean and extensible architecture,
\cpachecker is an ideal tool for teaching of program-analysis techniques.
The techniques can be explored in comparison and their effects observed.
Visualizations of abstract states and error paths help understand the reasons
for correctness or violation of the specification.
We use \cpachecker in various courses on software engineering, software verification,
software testing, and program semantics.

\subsection{Outline}

This tutorial starts in~\cref{sec:getting-started}
with installation instructions and a first example of running \cpachecker.
\Cref{sec:inputs} explains the inputs and outputs of \cpachecker.
Finally, \cref{sec:algorithms} gives an overview on the most important analysis techniques
that \cpachecker provides for software verification.

This extended version of~\cite{CPAchecker-3.0} includes the following 
further information on \cpachecker.
\Cref{sec:examples-list} provides an overview of all concrete example command lines
together with references to the respective part of the tutorial.
\Cref{appendix:projectInfo} 
provides more information about the \cpachecker project,
its development history, achievements, and licensing.
\Cref{appendix:analysisDetails} 
provides some more detailed examples for the presented analysis techniques.
\cref{appendix:witness-validation,appendix:test-gen}
explain how to use \cpachecker for witness validation and test-case generation, respectively.

\section{Getting Started with \cpachecker}
\label{sec:getting-started}
In the following, we explain the installation and a few alternatives for executing
\cpachecker on individual verification tasks.

For trying out \cpachecker and following this tutorial
we provide a few example programs
in a reproduction package~\cite{CPAcheckerTutorial-artifact-FM24-proceedings}.
We assume this package was downloaded and unpacked, and that the current working directory is its
root directory (where directory \texttt{examples/} is visible).
The execution of each example in this tutorial
should take less than 10~seconds.

\subsection{Local Installation}
\label{sec:installation}

\inlineheadingbf{Installation Requirements}
Most features of \cpachecker require a 64-bit GNU/Linux machine,
unless users build the required libraries themselves.
A limited feature set is usable on other platforms.
We recommend a current LTS version of Ubuntu;
recent versions of other distributions can be expected to work as well.

\inlineheadingbf{Installation}
For users on Debian or Ubuntu we provide a package repository
at \url{https://apt.sosy-lab.org}.
Please follow the instructions on that webpage to enable the repository.
Afterwards, the latest version of \cpachecker can be installed with
\texttt{sudo apt install cpachecker}.

For users without root access or on other distributions,
we also provide \cpachecker as pre-built binary releases via
\href{https://doi.org/10.5281/zenodo.3816620}{Zenodo}~\cite{CPAchecker-latest}
and our \href{https://cpachecker.sosy-lab.org/download.php}{download page}.
Please ensure that a Java Runtime Environment (JRE) is available
(for \cpachecker~3.0, Java version 17 or newer is required).
Unpack the archive for \cpachecker after the download.
We recommend adding \cpachecker's \texttt{bin/} directory
to the \texttt{PATH} environment variable.
This way the examples provided in this tutorial
work as is, without having to specify the full path to the \texttt{cpachecker} executable every time.
If \cpachecker was installed via the package repository, changing the \texttt{PATH} variable is not necessary.

\inlineheadingbf{Execution}
To try out \cpachecker,
run the following command from the reproduction package's~\cite{CPAcheckerTutorial-artifact-FM24-proceedings}
root directory:
\begin{verbatim}
cpachecker examples/example-safe.c
\end{verbatim}
This will verify that there is no assertion violation in program \exampleprog{example-safe.c},
and report that the program satisfies the specification.
Further information is provided in \cref{sec:hello-world}.

\subsection{Execution via Container}

\cpachecker is available as an image in OCI format, for use with
container runtimes like \href{https://podman.io/}{Podman}
and Docker.
The identifiers of the images are
{\smaller\href{https://hub.docker.com/r/sosylab/cpachecker/}{\texttt{sosylab/\allowbreak cpachecker}}}
(always the latest release)
and {\smaller\texttt{sosylab/\allowbreak cpachecker:3.0}} for version~3.0.
The following command line executes \cpachecker~3.0 from a container
(may require \texttt{sudo}, depending on the Docker installation):
\begin{verbatim}
docker run -v "$(pwd)":/workdir sosylab/cpachecker:3.0 \
    examples/example-safe.c
\end{verbatim}
Command-line argument \mbox{\texttt{-v "\$(pwd)":/workdir}}
makes the current working directory (\texttt{\$(pwd)})
available in the started container at path \texttt{/workdir}.
This is the default entrypoint of the \cpachecker images.
Command-line argument \mbox{\texttt{-u \$UID:\$GID}} might be added
after \texttt{docker run}
to set the user
and group ID of the container to the current user and group ID:
output files produced by \cpachecker are then owned by the current user instead of \texttt{root}.
Argument \texttt{examples/example-safe.c} is
passed to \cpachecker and will be explained in \cref{sec:hello-world}.
The command-line arguments and input files can be adjusted as usual.

\subsection{Remote Execution via Website}
We provide a web interface for \cpachecker
at \url{https://vcloud.sosy-lab.org/cpachecker/webclient/run/}.
The examples of this paper are available as Examples %
on the left of the page.

\subsection{Example Verification Task}
\label{sec:hello-world}

\begin{tcolorbox}
    For all example command lines in this paper we assume a local installation of \cpachecker
    and that the artifact with the examples~\cite{CPAcheckerTutorial-artifact-FM24-proceedings}
    has been unpacked in the current directory
    (such that the directory \texttt{examples/} is present).
    If necessary, e.g., for Docker usage, please adjust the command lines accordingly.
\end{tcolorbox}

\inlineheadingbf{Program Description}
We use the program in \cref{fig:ex-safe-c}.
This program initializes variables \texttt{n} and \texttt{x}
to two nondeterministic but concrete values of type \texttt{unsigned int} 
(modeled by calls to \texttt{__VERIFIER_nondet_uint()})
and then initializes \texttt{y} to the difference between \texttt{n} and \texttt{x}.
As long as~\texttt{x} is larger than \texttt{y},
the \texttt{while} loop decrements \texttt{x}
and increments \texttt{y} by one.
If the sum of \texttt{x} and \texttt{y} does not equal \texttt{n}
at the end of a loop iteration,
\texttt{__assert_fail} at \cref{ex-safe-err} triggers a program error
(arguments omitted for simplicity).
The program is correct with respect to the specification that
\texttt{__assert_fail} is unreachable,
because the sum of \texttt{x} and \texttt{y} always equals \texttt{n}
at the end of every loop iteration.
A variant of this program is shown in \cref{fig:ex-unsafe-c}.
The variant follows the same execution except at \cref{ex-unsafe-if}.
Here an error is triggered if \texttt{x} is smaller than~\texttt{y}.
This error is reachable by initializing \texttt{n} to~3 and \texttt{x} to~2
(among many other possibilities).

\newsavebox{\exampleSafeCode}
\begin{lrbox}{\exampleSafeCode}
    \begin{minipage}[b]{0.46\textwidth}
        \centering
\begin{lstlisting}[
            style=C,
            breaklines=true,
            %
        ]
extern unsigned __VERIFIER_nondet_uint();
extern void __assert_fail();
int main() {
  unsigned n =  __VERIFIER_nondet_uint();
  unsigned x =  __VERIFIER_nondet_uint();
  unsigned y = n - x;(*@\label{ex-safe-inity}@*)
  while(x > y) {(*@\label{ex-safe-while}@*)
    x--; y++;
    if ((*@\textcolor{green}{x + y != n}@*)) {(*@\label{ex-safe-if}@*)
      __assert_fail();(*@\label{ex-safe-err}@*)
    }
  }(*@\label{ex-safe-endwhile}@*)
  return 0;
}
\end{lstlisting}
    \end{minipage}
    \vspace{-3mm}
\end{lrbox}

\newsavebox{\exampleUnsafeCode}
\begin{lrbox}{\exampleUnsafeCode}
    \begin{minipage}[b]{0.46\textwidth}
        \centering
\begin{lstlisting}[
            style=C,
            breaklines=true,
            %
        ]
extern unsigned __VERIFIER_nondet_uint();
extern void __assert_fail();
int main() {
  unsigned n =  __VERIFIER_nondet_uint();(*@\label{ex-unsafe-initn}@*)
  unsigned x =  __VERIFIER_nondet_uint();(*@\label{ex-unsafe-initx}@*)
  unsigned y = n - x;(*@\label{ex-unsafe-inity}@*)
  while(x > y) {(*@\label{ex-unsafe-while}@*)
    x--; y++;
    if ((*@\textcolor{red}{x < y}@*)) {(*@\label{ex-unsafe-if}@*)
      __assert_fail();(*@\label{ex-unsafe-err}@*)
    }
  }
  return 0;
}
\end{lstlisting}
    \end{minipage}
\end{lrbox}

\begin{figure}[t]
    \centering
    \subfloat[\exampleprog{example-safe.c} (error unreachable)]{\usebox{\exampleSafeCode}\label{fig:ex-safe-c}}
    \hfill
    \subfloat[\exampleprog{example-unsafe.c} (error reachable)]{\usebox{\exampleUnsafeCode}\label{fig:ex-unsafe-c}}
    \caption{Example C programs}
    \label{fig:example-code}
    \vspace{-4mm}
\end{figure}

\inlineheadingbf{Verification Run}
To verify the example program in \cref{fig:ex-safe-c}
with \cpachecker,
execute the below command in a terminal
 \exonline{default}:
\begin{verbatim}
cpachecker examples/example-safe.c
\end{verbatim}
This command line does not specify an explicit configuration.
In this case \cpachecker uses the default configuration,
which is the currently recommended configuration.
Like most configurations shipped with \cpachecker,
the default configuration uses the default specification,
which specifies that no C~assertion error \texttt{__assert_fail}
and no label named \texttt{ERROR}
should be reachable.
The specifications, configurations,
and the available analyses
are described in more detail in \cref{sec:specification}, \cref{sec:config}, and \cref{sec:algorithms}.

\begin{samepage}
At the end of its execution, \cpachecker produces the following messages:
  \begin{Verbatim}[breaklines=true,fontsize=\fontsize{8}{9}]
Verification result: TRUE. No property violation found by chosen configuration.
More details about the verification run can be found in the directory "./output".
Graphical representation included in the file "./output/Report.html".    
  \end{Verbatim}
\end{samepage}
The verification result \texttt{TRUE} indicates that the error (\cref{ex-safe-err} in \cref{fig:ex-safe-c})
is not reachable.
We can also change the input program to \exampleprog{example-unsafe.c} in the command line.
In this case, the verification result is \texttt{FALSE},
meaning that \cpachecker finds an execution path that triggers the error.
The meanings of verification results and how to navigate through the generated report
is the topic of \cref{sec:verdict} and \cref{sec:report}, respectively.

\section{Input and Output Interface of \cpachecker}
\newcommand{\symWitness}{\omega}
\newcommand{\symProgram}{P}
\newcommand{\symSpecification}{\varphi}
\newcommand{\symVerdict}{v}
\newcommand{\symConfiguration}{\rho}
\newcommand{\symReport}{R}

\label{sec:inputs}
\Cref{fig:inputsOutputs} shows the inputs and outputs of \cpachecker.
\cpachecker always takes a program, a specification,
and a configuration
as input.
It always produces
a verdict and
a report.
Depending on how the user intends to use it,
either as a verifier, a witness validator, or a test-case generator,
\cpachecker may also take a verification witness as input,
or produce witnesses or test cases as output.

\subsection{Input Program}\label{sec:programs}
\cpachecker supports a large subset of the GNU-C11 features.
Normally, the verifier expects pre-processed input files.
\cpachecker supports compiler directives
(e.g., \texttt{\#include} or \texttt{\#define})
if the command-line argument \texttt{-{}-preprocess} is given,
in which case \cpachecker pre-processes the input C program.
To guarantee a meaningful verification of programs that use external functions,
including functions in the C standard library,
the implementations of the functions have to be provided in the input programs.
Otherwise, \cpachecker overapproximates their behavior, potentially leading to false alarms.
Two exceptions are the function \texttt{pthread_create} for creating a new thread
and functions \texttt{malloc}, \texttt{memset}, etc., for manipulating memory,
which are handled out-of-the-box by \cpachecker's concurrency and memory analyses, respectively.
To verify a software project that consists of multiple C files,
all relevant files must be listed on the command-line.
By default, \cpachecker starts the analysis from the function \texttt{main}.
Another entry function can be
specified with the command-line argument \texttt{-{}-entry-function <entry function>}.

The semantics of a C~program depends on the runtime platform,
which consists of a machine architecture, a data model, and an operating system.
\cpachecker assumes a single platform during verification.
The command-line argument~\texttt{-{}-32} (default)
sets the platform to 32-bit x86 Linux (ILP32) and
\texttt{-{}-64} sets the platform to 64-bit x86 Linux (LP64)~\cite{64bit}.

\subsection{Program Specification}\label{sec:specification}
Besides the input program, a \emph{specification} is needed as input for \cpachecker.
The specification defines what property of the program should be checked.
\cpachecker supports an automaton-based specification language
(similar to \blast~\cite{BLAST-query} and \slam~\cite{SLIC})
to define program specifications
(documented in \href{https://svn.sosy-lab.org/software/cpachecker/tags/cpachecker-3.0/doc/SpecificationAutomata.md}{\texttt{doc/SpecificationAutomata.md}}).
\cpachecker ships with several common specifications
in the directory \href{https://svn.sosy-lab.org/software/cpachecker/tags/cpachecker-3.0/config/specification}{\texttt{config/specification/}}.
A selection is listed in \cref{tab:specifications}.
\cpachecker also supports
\href{https://gitlab.com/sosy-lab/benchmarking/sv-benchmarks/-/tree/main/c/properties}{property files}
written in the specification language that was standardized by the International Competition on Software Verification (\svcomp)~\cite{SVCOMP24}.

\begin{table}[t]
	\centering
	\caption{Provided specifications
  (files in \href{https://svn.sosy-lab.org/software/cpachecker/tags/cpachecker-3.0/config/specification}{\texttt{config/specification/}})}
	\label{tab:specifications}
	\rowcolors{0}{}{black!10}
	\begin{tabular}{lp{.80\textwidth}}
		\toprule
		Specification                                    & Description
		\\
		\midrule
		\cpaspec{ErrorLabel}
			& Labels named \texttt{ERROR} (case insensitive) are never reachable.
		\\
		\cpaspec{Assertion}
			& All \texttt{assert} statements hold.
		\\
		\cpaspec{default}
			& Both \cpaspec{ErrorLabel} and \cpaspec{Assertion} hold.
		\\
		\cpaspec{overflow}
			& All operations with a signed-integer type never produce values outside the range representable by the respective type.
		\\
		\cpaspec{datarace}
			& Concurrent accesses to the same memory location must be atomic if at least one of them is a write access.
		\\
		\cpaspec{memorysafety}
			& All memory deallocations and pointer dereferences are valid and all allocated memory is pointed to or deallocated when the program exits.
		\\
		\cpaspec{memorycleanup}
			& All allocated memory is deallocated before the program exits.
		\\
		\bottomrule
	\end{tabular}
    \vspace{-3mm}
\end{table}

The command-line argument \texttt{-{}-spec <specification>}
defines the specification to use.
It accepts the path to a specification-automaton file, an \svcomp property file,
or the name of one of the specifications that ship with \cpachecker.
For example, to verify a program against the provided specification \cpaspec{Assertion}
with \cpachecker's default analysis, we run \exonline{assert}:
\begin{verbatim}
cpachecker [--preprocess] --spec Assertion examples/example-safe.c
\end{verbatim}
The square brackets in the above command indicate that
argument \texttt{-{}-preprocess} may be omitted if the program
does not contain compiler directives (cf. \cref{{sec:programs}}).

\Cref{fig:bql} shows a simplified version of the \texttt{Assertion} specification.
The specification is violated if a call to function \texttt{__assert_fail}
is reachable in the given input program,
which matches how \texttt{assert} statements appear in a C~program after pre-processing.
The automaton
starts in the initial state~\texttt{Init}
and observes the analyzed program operations
until an operation matches a call to \texttt{__assert_fail} (\cref{spc:match})
with an arbitrary number of function-call arguments (denoted by~\texttt{\$?}).
In this case,
the automaton transitions to the special state~\texttt{ERROR}~(\cref{spc:error})
that signals a specification violation with the given explanation.

\begin{figure}[t]
\begin{lstlisting}[style=spc,escapechar=&]
OBSERVER AUTOMATON AssertionErrorAutomaton&\label{spc:automaton-start}&
INITIAL STATE Init;&\label{spc:init}&
STATE USEFIRST Init :&\label{spc:usefirst}&
  // AST-based matching of function calls to __assert_fail
  MATCH {__assert_fail($?)}&\label{spc:match}&
    -> ERROR("assertion in $location");&\label{spc:error}&
END AUTOMATON&\label{spc:automaton-end}&
\end{lstlisting}
    \vspace{-3mm}
	\caption{Example of automaton-based specification for checking assert statements}
	\label{fig:bql}
    \vspace{-5mm}
\end{figure}


\subsection{\cpachecker Configuration}
\label{sec:config}
\cpachecker is highly configurable via a set of configuration options,
which are documented in the file
\href{https://svn.sosy-lab.org/software/cpachecker/tags/cpachecker-3.0/doc/ConfigurationOptions.txt}{\texttt{doc/ConfigurationOptions.txt}}.
Configuration options are specified as key-value pairs in a configuration file
or on the command line.
An extensive set of bundled configuration files is available in directory
\href{https://svn.sosy-lab.org/software/cpachecker/tags/cpachecker-3.0/config}{\texttt{config/}}.
Most of these bundled configurations
specify default values for common configuration options,
e.g., the specification \href{https://svn.sosy-lab.org/software/cpachecker/tags/cpachecker-3.0/config/specification/default.spc}{\texttt{config/specification/default.spc}}
and a time limit of~\SI{900}{\second}.
Command-line arguments overwrite these defaults.

It is possible to write and provide own configuration files.
Their format is inspired by Windows INI files with some extensions like include directives.
A~full description is available in
\href{https://svn.sosy-lab.org/software/cpachecker/tags/cpachecker-3.0/doc/Configuration.md#configuration-file-format}{\texttt{doc/Configuration.md}}.
Configuration files may use relative paths.
\cpachecker interprets these relative paths
relative to the directory of the respective configuration file.

Command-line argument \texttt{-{}-config CONFIG_FILE}
selects a configuration file.
The bundled configuration files can also be selected with short-hand arguments that
consist of the base name of the configuration file,
e.g., \texttt{-{}-kInduction} for the configuration file
\href{https://svn.sosy-lab.org/software/cpachecker/tags/cpachecker-3.0/config/kInduction.properties}{\texttt{config/kInduction.properties}}
or \texttt{-{}-svcomp24} for the configuration file
\href{https://svn.sosy-lab.org/software/cpachecker/tags/cpachecker-3.0/config/svcomp24.properties}{\texttt{config/svcomp24.properties}}.
When no configuration file is explicitly specified,
\cpachecker runs in its default configuration
(defined by the configuration file
\href{https://svn.sosy-lab.org/software/cpachecker/tags/cpachecker-3.0/config/default.properties}{\texttt{config/default.properties}}).

The command-line argument \texttt{-{}-option key=value}
sets a single configuration option.
The order of command-line arguments is irrelevant.
If an option is set both in the configuration file and
through \texttt{-{}-option}, the \texttt{-{}-option} value takes precedence
and overwrites any value from the configuration file.

\cpachecker provides shortcuts for the most common configuration options,
for example \texttt{-{}-64} to specify the platform as 64-bit x86 Linux (LP64),
or \texttt{-{}-timelimit} to set an analysis time limit.
A full list of shortcuts is available via \texttt{cpachecker -h}
and in \href{https://svn.sosy-lab.org/software/cpachecker/tags/cpachecker-3.0/doc/Configuration.md}{\texttt{doc/Configuration.md}}.
For technical reasons, a few command-line arguments exist
that can only be specified through command-line arguments and not via configuration files.
These arguments include
\texttt{-{}-benchmark} (which leads to better performance
by disabling \cpachecker-internal assertions, writing no output files, and much more)
and
\texttt{-{}-heap} (which adjusts the amount of memory used by the JVM for \cpachecker).

As an example, consider the following command line \exonline{settingOptions}:
\begin{verbatim}
cpachecker --kInduction --timelimit 900s --heap 2000M \
    --spec ErrorLabel examples/example-safe.c \
    --option solver.solver=MATHSAT5
\end{verbatim}
This invokes \cpachecker with the configuration for \kinduction,
sets the configuration option \texttt{limits.time.cpu} for the time limit to~\SI{900}{\second},
tells the JVM to use \SI{2000}{\mebi\byte} of heap memory,
chooses the specification file \cpaspec{ErrorLabel},
the program \texttt{program.c} as input file,
and sets the configuration option \texttt{solver.solver} to \texttt{MATHSAT5}.

\subsection{Verification Verdict}
\label{sec:verdict}

\cpachecker may report three different verification verdicts:
(1)~\texttt{TRUE}, if it proves that the program \emph{satisfies} the specification;
(2)~\texttt{FALSE}, if it proves that the program does \emph{not satisfy} the specification;
(3)~\texttt{UNKNOWN}, if it cannot decide the verification task using the given resource limits and configuration.

\subsection{Interactive Report in HTML Format}
\label{sec:report}

In addition to a verification verdict,
\cpachecker produces detailed information about the performed analysis in directory \texttt{output/}
in the current working directory.
This usually includes an interactive report in HTML format.
Note that different configurations may produce different output files.

\begin{figure}[t]
  \includegraphics[width=\textwidth]{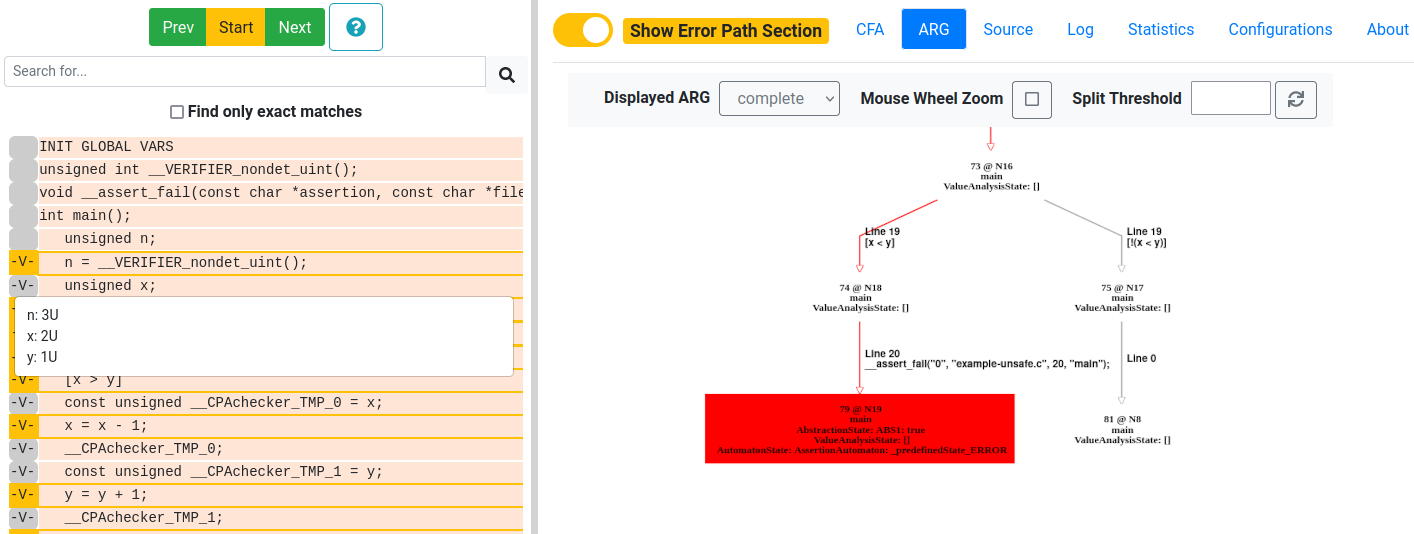}
  \caption{Screenshot of the HTML report for program \exampleprog{example-unsafe.c}}
  \label{fig:report}
  \vspace{-5mm}
\end{figure}

The interactive report
offers a graphical interface for users to inspect the results of \cpachecker.
It allows to inspect, among others:
the \textit{control-flow automata}~(CFA) of the input program,
the \textit{abstract reachability graph} (ARG) that was constructed by the chosen configuration,
statistics,
and an error path that violates the specification
(if the verdict is \texttt{FALSE}).

In the following we explain the most important parts of this report.
A~screenshot of the report is shown in~\cref{fig:report}.
\href{https://cpachecker.sosy-lab.org/counterexample-report/ErrorPath.0.html}{An example report} is provided online.
If \cpachecker reports the verdict \texttt{FALSE}
and the used analysis provides detailed counterexample information,
the report file is \texttt{output/\allowbreak Counterexample.0.html}
(number \texttt{0} may differ).
Otherwise, the report file is \texttt{output/\allowbreak Report.html}.

\inlineheadingbf{Control-Flow Automata}
The tab \reporttab{CFA} in the report shows the input program in
the internal representation of \cpachecker,
the control-flow automata~(CFA).
A CFA consists of program locations (nodes of the graph)
and program statements (edges of the graph).
In the report, a double-click on a CFA edge
navigates to the source-code line it represents.
The drop-down menu ``Displayed CFA'' can be used to
display a single CFA for a single program function.

\inlineheadingbf{Abstract-Reachability Graph}
The tab \reporttab{ARG} in the report shows a graphical representation
of the program states that were explored by \cpachecker
in the form of an abstract-reachability graph~(ARG).
The right-hand side of \cref{fig:report} shows an ARG.
Each node in the ARG represents an \textit{abstract} state of the input program.
\cpachecker constructs abstract states according to the selected configuration.
An abstract state usually represents a set of \textit{concrete} program states
in order to overapproximate the reachable state space.
Two abstract states are connected by a directed edge
if one state is the successor to the other.
The directed edge goes from predecessor to successor and
is labeled with a program operation that induced the predecessor-successor relation
during analysis.

If \cpachecker reported the verdict \texttt{TRUE},
the ARG represents all reachable abstract program states.
If \cpachecker reported the verdict \texttt{FALSE},
nodes and edges that are part of the error path are marked in red (as in \cref{fig:report}).

\inlineheadingbf{Error Path}
If the verification verdict is \texttt{FALSE}
and the analysis provides detailed counterexample information,
the report includes a textual error-path section as separate panel on the left
(toggle with button ``Show Error-Path Section'').
This allows users to step through the error path that \cpachecker computed.
The textual error path is a list of program statements,
accompanied by concrete assignments to all variables on the error path.
A button \inlinebutton{-V-}{vButtonColor} is displayed next to each statement,
which indicates the concrete variable assignments at the respective location.
To replay the error path step-by-step,
users can click on the
\inlinebutton{Start}{startButtonColor} button on the top left.
Then, two buttons \inlinebutton[white]{Next}{prevNextButtonColor}
and \inlinebutton[white]{Prev}{prevNextButtonColor} can be used to navigate through the error path.

\subsection{Statistics}

\cpachecker collects a variety of statistics, depending on the chosen analysis.
These are presented in the interactive report under tab \reporttab{Statistics}
and are also written to file \texttt{output/Statistics.txt}.
With the command-line argument \texttt{-{}-stats},
\cpachecker prints the statistics to the console at the end of the verification run.

The statistics help users to evaluate the performance of the analysis.
Below is an example excerpt of a run's statistics that shows
the time spent on SMT~solving,
the total number of computed reachable abstract states,
and the consumed CPU time.
\begin{Verbatim}
Total time for SMT solver (w/o itp):     0.017s
[...]
Size of reached set:             10
[...]
CPU time for analysis:            0.860s
\end{Verbatim}
\href{https://svn.sosy-lab.org/software/cpachecker/tags/cpachecker-3.0/doc/tutorials/interpret-statistics.md}{A separate tutorial}
covers how to interpret \cpachecker statistics in more detail.

\subsection{Verification Witnesses}
\label{sec:output-witness}

Verification witnesses~\cite{WitnessesJournal,VerificationWitnesses-2.0} help users and tools to reason about verification results
and allow independent validation of the verification result.
Validation is usually easier than verification, thanks to
the additional information the witness provides.
\cpachecker can both export witnesses for verification results
and validate witnesses that other tools produce.
\Cref{appendix:witnesses} explains witness validation in detail.

\newsavebox{\exampleSafeWitnessVB}
\begin{lrbox}{\exampleSafeWitnessVB}
\begin{lstlisting}[style=yaml,numbers=none]





  <...>
  content:
  - invariant:
      type: "loop_invariant"
      location:
        file_name: "example-safe.c"
        line: (*@\ref{ex-safe-while}@*)
        column: 3
        function: "main"
      value: "( x + y == n )"
      format: "c_expression"
\end{lstlisting}
\end{lrbox}

\newsavebox{\exampleUnsafeWitnessVB}
\begin{lrbox}{\exampleUnsafeWitnessVB}
\begin{lstlisting}[style=yaml,numbers=none]
<...>
content:
 - segment:
   - waypoint:
       type: assumption
       location:
         file_name: "example-unsafe.c"
         line: (*@\ref{ex-unsafe-inity}@*)
       constraint:
         value: "x == 0 && n == 1"
 - segment:
   - waypoint:
       type: target
       location:
         file_name: "example-unsafe.c"
         line: (*@\ref{ex-unsafe-err}@*)
\end{lstlisting}
\end{lrbox}

\begin{figure}[t]
    \subfloat[Relevant sections of a correctness witness for the safe program in~\cref{fig:ex-safe-c}]
      {\usebox{\exampleSafeWitnessVB}\label{fig:witnesses:correctness-witnessesv2}}
    \hfill
    \subfloat[Relevant sections of a violation \\ witness for the unsafe program in~\cref{fig:ex-unsafe-c}]
      {\usebox{\exampleUnsafeWitnessVB}\label{fig:witnesses:violation-witnessesv2}}
    \vspace{-1.5mm}
    \caption{Example verification witnesses (format version 2.0, slightly shortened for readability)}
    \label{fig:witnesses:witnessesv2}
    \vspace{-5mm}
\end{figure}

\inlineheadingbf{Correctness Witnesses}
Correctness witnesses are defined for reachability of error locations
and detection of signed-integer overflows
in sequential programs.
\cpachecker produces such a witness
not only if the verdict is \texttt{TRUE},
but also if it is \texttt{UNKNOWN}
(in this case with partial information).
The witness contains information about the explored program state space
in the form of loop and location invariants.
In case the analysis result is \texttt{TRUE}, the invariants hold
whenever the program execution passes through the respective location.

\Cref{fig:witnesses:correctness-witnessesv2}
shows an excerpt of a correctness witness for the safe program in \cref{fig:ex-safe-c}.
It reports the loop invariant \texttt{x + y == n}
for the loop head in \cref{ex-safe-while}.

\inlineheadingbf{Violation Witnesses}
Violation witnesses represent one or more program paths that lead
to a specification violation.
This is achieved by specifying assumptions about the program inputs and the control flow of the program.

\Cref{fig:witnesses:violation-witnessesv2}
shows an excerpt of a violation witness for the unsafe program in \cref{fig:ex-unsafe-c}.
It shows the program path that leads to the assertion failure at \cref{ex-unsafe-err}
when \texttt{x} is assigned value 0 and \texttt{n} is assigned value 1.

\subsection{Test Harnesses}
\label{sec:output-tests}
If \cpachecker
finds a specification violation (verdict~\texttt{FALSE}),
it produces a test harness that triggers this violation through test execution.
A test harness contains a sequence of external inputs
(e.g., for inputs modeled by \texttt{__VERIFIER_nondet*}) to the program
that trigger an execution path to the specification violation.
\Cref{fig:test-harness} shows an excerpt of a test harness for the example program in \cref{fig:ex-unsafe-c}.
The two return values \texttt{2U} (lines~5 and~6)
initialize, in the program under analysis (\cref{fig:ex-unsafe-c}),
both variables \texttt{n} and \texttt{x} with value~2.
This triggers the assertion failure at \cref{ex-unsafe-err} of the program.

\newsavebox{\exampleTestHarness}
\begin{lrbox}{\exampleTestHarness}
  \centering
  \begin{lstlisting}[style=C]
unsigned __VERIFIER_nondet_uint() {
  static unsigned call_count = 0;
  unsigned retval;
  switch (call_count) {
    case 0: retval = 2U; break;
    case 1: retval = 2U; break;
  }
  ++call_count;
  return retval;
}
  \end{lstlisting}
\end{lrbox}
\begin{figure}[t]
  \hspace{6.5mm}\usebox{\exampleTestHarness}
  \vspace{-3mm}
  \caption{Test harness generated for the example program in \cref{fig:ex-unsafe-c}}
  \label{fig:test-harness}
  \vspace{-5mm}
\end{figure}

The test harness can be compiled
with the program under analysis:
\begin{Verbatim}
gcc output/Counterexample.1.harness.c examples/example-unsafe.c
\end{Verbatim}
This produces a binary \texttt{a.out}.
The execution of \texttt{./a.out}
exhibits that the claimed specification violation is actually reachable.
It reports:
\begin{Verbatim}
CPAchecker test harness: property violation reached
\end{Verbatim}
\Cref{appendix:test-gen} gives more details on test generation with \cpachecker.

\section{Verification Analyses and How to Select Them}\label{sec:algorithms}

This section shows how to execute various commonly-used verification analyses in \cpachecker.
These analyses can be divided into three groups depending
on the kind of specifications they can check.
First, there are analyses that perform a reachability analysis.
These support common specifications, for example,
reachability of an error location or an assertion violation.
Second, there are analyses that support a particular special-purpose specification.
Third, there are meta analyses that implement strategy selection
and delegate to one of the above depending on the provided specification.
\Cref{tab:recommended-analyses} lists common
configurations and the respective specifications they support.
Apart from the configuration \cpaconfig{dataRaceAnalysis},
which performs partial order reduction~\cite{POR98} over memory accesses
in combination with value analysis~\cite{CPAexplicit},
the following sections explain these configurations in more detail.

\begin{table}[t]
  \caption{Commonly-used configurations and supported specifications}
  \label{tab:recommended-analyses}
  \centering
  \begin{tabular}{!{\hspace{3mm}}l@{\hspace{1mm}}p{5.2cm}<{\raggedright}l}
    \toprule
    Configuration                             & Specification (cf. \cref{sec:specification})                                                                                                                                 & Description                             \\
    \midrule
    \multicolumn{3}{l}{Configurations for reachability specifications:}                                                                                                                                                                                       \\
    \cpaconfig{valueAnalysis-NoCegar-join}    & \multirow{5}{5.5cm}{\cpaspec{default}, \cpaspec{Assertion}, \cpaspec{ErrorLabel}, custom automaton specifications,\linebreak and \svcomp property \href{https://gitlab.com/sosy-lab/benchmarking/sv-benchmarks/-/tree/main/c/properties/unreach-call.prp}{\texttt{unreach-call.prp}}} & \Cref{sec:val}                          \\
    \cpaconfig{symbolicExecution-NoCegar}     &                                                                                                                                                                              & \Cref{sec:symex}                        \\
    \cpaconfig{predicateAnalysis}             &                                                                                                                                                                              & \Cref{sec:predabs}                      \\
    \cpaconfig{bmc-incremental}               &                                                                                                                                                                              & \Cref{sec:bmc}                          \\
    \cpaconfig{kInduction}                    &                                                                                                                                                                              & \Cref{sec:bmc-extensions}               \\
    \midrule
    \multicolumn{3}{l}{Special-purpose configurations:}                                                                                                                                                                                                                \\
    \cpaconfig{smg}                           & memory safety (\cpaspec{memorysafety}\linebreak and \cpaspec{memorycleanup})                                                                                                 & \Cref{sec:smg}                          \\
    \cpaconfig{lassoRankerAnalysis}           & \multirow{2}{*}{termination}
                                                                                                                             & \multirow{2}{*}{\Cref{sec:termination}} \\
    \cpaconfig{terminationToSafety}          &                                                                                                                                                                              &                                         \\
    \cpaconfig{predicateAnalysis--overflow} & \cpaspec{overflow}                                                                                                                                                           & \Cref{sec:overflow}                     \\
    \cpaconfig{dataRaceAnalysis}              & \cpaspec{datarace}                                                                                                                                                           & \Cref{sec:algorithms}                   \\
    \midrule
    \multicolumn{3}{l}{Meta configurations:}                                                                                                                                                                                                                           \\
    \cpaconfig{svcomp24}                      & \multirow{2}{5.2cm}{reachability specifications\linebreak and all \href{https://gitlab.com/sosy-lab/benchmarking/sv-benchmarks/-/tree/svcomp24/c/properties}{\svcomp properties}}                                                                               &\!\!\cite{CPACHECKER-SVCOMP24}              \\
    default (no argument)                     &                                                                                                                                                                              & \Cref{sec:analysis-selection}                  \\
    \bottomrule
  \end{tabular}
  \vspace{-5mm}
\end{table}

\subsection{Selecting an Analysis}
\label{sec:analysis-selection}

Selecting an analysis of \cpachecker primarily depends on the kind of specification
that should be verified.
Memory safety, overflows, and data races
can each be verified by exactly one recommended analysis,
which is listed in \cref{tab:recommended-analyses}.
For termination, there are two recommendations, described in \cref{sec:termination}.
If SV-COMP property files are used to encode the specification,
meta configurations of \cpachecker automatically select
a recommended analysis depending on the specification.

For standard reachability specifications
a wide range of different analyses and techniques
is available in \cpachecker.
Each of them has their strengths and weaknesses,
and while some of them are more powerful or efficient in general,
none of them always outperforms all of the others,
so it can be worthwhile experimenting with several analyses.

The general recommendation for most use cases
is the default analysis of \cpachecker
(used if no other configuration is selected on the command line).
It is a meta configuration that uses \kinduction
(\cpaconfig{kInduction}, most effective overall in our experience)
for reachability specifications.

\cpachecker's value analysis (\cpaconfig{valueAnalysis-NoCegar-join}),
symbolic execution (\cpaconfig{symbolicExecution-NoCegar}),
and bounded model checking (BMC, \cpaconfig{bmc-incremental})
are mostly suited for finding specification violations.
While they are often quite efficient in finding bugs,
they are often inefficient for proving correctness for large programs.
In our experience these configurations usually either succeed quickly
or will not produce a result at all.

To prove the absence of specification violations in larger programs,
either abstraction of the program state space
or a proof technique such as induction needs to be used.
Value analysis and symbolic execution
support a limited form of abstraction
(ignoring irrelevant program variables and clauses)
if their configuration variants with precision refinement are chosen
as described in the respective sections below.
Predicate abstraction (\cpaconfig{predicateAnalysis}) is stronger
and can in principle find arbitrary loop invariants
as long as the loop invariants do not require quantifiers nor floating-point arithmetic.
\kInduction (\cpaconfig{kInduction}) on the other hand
requires that an induction proof can be found for the program.

Another aspect that needs to be considered
is that value analysis and symbolic execution in \cpachecker
do not support precise reasoning about dynamically allocated memory
and data structures on the heap,
whereas BMC, predicate abstraction, and \kinduction do support this.
However, the latter three are based on solving (sometimes large)
formulas with an SMT solver, which may not scale.
Value analysis has the advantage that it does not require SMT solving,
but the disadvantage that it cannot reason about non-deterministic values.
Symbolic execution uses an SMT solver,
but only when required for non-deterministic values.

The value analysis can be considered
comparatively easy to understand conceptually,
which makes it a good starting point for the use of \cpachecker.

\subsection{Value Analysis}
\label{sec:val}

\begin{table}[t]
  \caption{Main configuration flavors of value analysis}
  \label{tab:valueConfigs}
  \centering
  \begin{tabular}{c@{\hspace{3mm}}c@{\hspace{3mm}}l}
    \toprule
    \multicolumn{1}{l@{\hspace{3mm}}}{Precision Refinement} & \multicolumn{1}{l@{\hspace{3mm}}}{Path Sensitivity} & Configuration                          \\
    \midrule
    \xmark                                                  & \xmark                                              & \cpaconfig{valueAnalysis-NoCegar-join} \\
    \xmark                                                  & \cmark                                              & \cpaconfig{valueAnalysis-NoCegar}      \\
    \cmark                                                  & \cmark                                              & \cpaconfig{valueAnalysis-Cegar}        \\
    \bottomrule
  \end{tabular}
  \vspace{-2mm}
\end{table}

\cpachecker's value analysis tracks concrete value assignments.
There are two main configuration choices for the value analysis:
(1)~whether to use precision refinement, and
(2)~whether to be path sensitive.
\Cref{tab:valueConfigs} lists the available command-line arguments
to run \cpachecker with the corresponding configuration of value analysis.
For example, the following command runs a configuration of value analysis that
implements constant propagation~\cite{DragonBook} (no precision refinement, no path sensitivity)
on the program in \cref{fig:ex-const} \exonline{valueAnalysis-NoCegar-join}:
\begin{verbatim}
cpachecker --valueAnalysis-NoCegar-join examples/example-const.c
\end{verbatim}

\begin{figure}[t]
  \parbox[t]{.38\linewidth}{%
    \let\oldlabel\label
    \renewcommand\label[1]{\oldlabel{#1-a}}
\begin{lstlisting}[style=C]
extern void __assert_fail();
int main() {
  int x = 0;
  int y = 0;
  int z = 0;(*@\label{ex-const-initz}@*)
  while (x < 2) {(*@\label{ex-const-while}@*)
    x++;(*@\label{ex-const-xpp}@*)
    y = z + 1;(*@\label{ex-const-yz1}@*)
  }
  if (z != 0) {(*@\label{ex-const-if}@*)
    __assert_fail();
  }
  return 0;
}
\end{lstlisting}
  }\hfill%
  \parbox[t]{.58\linewidth}{%
    \let\oldlabel\label
    \renewcommand\label[1]{\oldlabel{#1-a}}
\begin{lstlisting}[style=C]
extern unsigned __VERIFIER_nondet_uint();
extern void __assert_fail();
int main() {
  unsigned int x = __VERIFIER_nondet_uint();
  unsigned int y = x;
  unsigned int z = __VERIFIER_nondet_uint();
  while (x < 2) {(*@\label{ex-symex-while}@*)
    x++;
    y++;
    z = x + z;
  }
  if (x != y) {(*@\label{ex-symex-if}@*)
    __assert_fail();
  }
  return 0;
}
\end{lstlisting}
  }\\
  \parbox[b]{.42\linewidth}{%
    \caption{Program \exampleprog{example-const.c}}
    \label{fig:ex-const}
  }\hfill%
  \parbox[b]{.58\linewidth}{%
    \caption{Program \exampleprog{example-sym.c}}
    \label{fig:ex-symex}
  }%
  \vspace{-6mm}
\end{figure}

This configuration tracks only value assignments
that always hold on a given location,
because abstract states are joined when control flow meets.
This is efficient, but in most cases not powerful enough to verify programs.
For~\cref{fig:ex-const}, it suffices because only the value
of variable~\texttt{z} is needed to prove the program safe,
and this is always~0.
\Cref{appendix:val} shows the state-space exploration of the value analysis
for this example in more detail.
If, however, the program safety would also depend
on the values of~\texttt{x} or~\texttt{y} after the loop,
the verification result would be \texttt{UNKNOWN}
because the analysis does not track these non-constant variable values.

The value analysis with path sensitivity tracks value assignments
per program path and location.
For the example in \cref{fig:ex-const}, it would keep track of all variable values
and fully unroll the loop.
This leads to path explosion
when many paths with distinct value assignments exist,
because the analysis tracks all of them separately.

Value analysis with path sensitivity and precision refinement
mitigates this path explosion by tracking
only those value assignments that are necessary for the analysis to prove the program safe.
This is more efficient than value analysis without precision refinement in the common case
where not all variables in the program are relevant for safety,
like in~\cref{fig:ex-const}.
The relevant variables are detected automatically
through counterexample-guided abstraction refinement (CEGAR)
with Craig interpolation~\cite{CPAexplicit}.

Because the value analysis always tracks concrete value assignments
and overapproximates nondeterministic values,
it may find false alarms. To mitigate this, \cpachecker
runs a precise, SMT-based feasibility check on every found potential error path
and only reports confirmed specification violations.
This can be seen in the output of \cpachecker, 
as shown in \cref{appendix:value-analysis-cex-check}.

\subsection{Interval-Based Data-Flow Analysis}
\label{sec:df}

The data-flow analysis (DF) of \cpachecker
is a lightweight proof-finding technique
that uses \emph{arithmetic expressions over intervals}
as its abstract domain~\cite{CPA-DF,kInduction-TR}.
It tracks, for an automatically-selected set of program variables,
the range of values that each variable can take
in the form of interval expressions,
e.g., $[l_1, u_1] \cup [l_2, u_2]$,
where $l_i$ (resp. $u_i$) is a numerical value
representing the lower (resp. upper) bound of an interval.
DF supports dynamic precision refinement.
At the beginning of the analysis,
it performs a coarse but efficient program exploration.
If some abstract state reachable in the exploration violates the safety specification,
DF incrementally increases its precision by
tracking more program variables,
allowing more complex expressions of intervals,
and disabling widening~\cite{AbstractInterpretation}.
To run DF in \cpachecker,
provide the configuration \cpaconfig{dataFlowAnalysis} on the command line \exonline{dataFlowAnalysis}:
\begin{verbatim}
cpachecker --dataFlowAnalysis examples/example-const.c
\end{verbatim}
For the above example, \cpachecker produces the verdict \texttt{TRUE}.
A limitation of DF is that
its abstract program exploration
cannot identify concrete error paths when there are specification violations
and may sometimes be too imprecise to find a safety proof.
For example, when \cpachecker analyzes \exampleprog{example-safe.c}
or \exampleprog{example-unsafe.c} in \cref{fig:example-code} with DF,
it produces the verdict \texttt{UNKNOWN}.
DF cannot only run standalone but also serve as an auxiliary invariant generator
that assists other analyses, e.g., \kinduction~\cite{kInduction}
(cf.~\cref{sec:bmc-extensions}).

\subsection{Symbolic Execution}
\label{sec:symex}
The symbolic execution~\cite{CPAsymexec-tool} of \cpachecker
tracks concrete value assignments
the same way as the value analysis.
But for every value that cannot be tracked concretely,
for example because it is assigned non-deterministically,
symbolic execution introduces a new symbolic value~$s_i$.
Whenever a symbolic value is used in an expression, symbolic execution
stores the expression over this symbolic value without evaluating it.
In addition, symbolic execution tracks the constraints over these symbolic values
for each program path.
This produces a symbolic-execution tree~(cf.\ \cref{appendix:symex} for details).
From this, concrete variable assignments can be derived for any program path.
The symbolic execution of \cpachecker also supports precision refinement
through \cegar with Craig interpolation~\cite{CPAsymexec}.
This determines which variables and constraints must be tracked
through the program.

The below command runs a configuration of symbolic execution~\cite{King76}
without precision refinement \exonline{symbolicExecution-NoCegar}:
\begin{verbatim}
cpachecker --symbolicExecution-NoCegar examples/example-sym.c
\end{verbatim}

Because symbolic execution tracks the expressions over symbolic values
without further abstraction, it is well suited for collecting constraints on inputs
for certain program paths.
But this precision also leads to path explosion:
The analysis of symbolic execution on program \exampleprog{example-safe.c} (\cref{fig:ex-safe-c})
does not terminate.
To prove the program safe, it is important to know that
the sum of~$x$ and~$y$ equals~$n$ at \cref{ex-safe-if}.
Symbolic execution tracks this by storing the expressions
$n = s_1, x = s_2, y = s_1 - s_2$,
$x = s_2 - 1, y = s_1 - s_2 + 1$,
$x = s_2 - 1 - 1, y = s_1 - s_2 + 1 + 1$, and so on.
This produces ever more complicated expressions and does not scale.

The following command runs a configuration of symbolic execution
with precision refinement \exonline{symbolicExecution-Cegar}:
\begin{verbatim}
cpachecker --symbolicExecution-Cegar examples/example-sym.c
\end{verbatim}

On the program of \cref{fig:ex-symex}, this only tracks assignments
and constraints over $x$ and $y$, which are necessary to prove the program safe.
Assignments to $z$ are not tracked.

\subsection{Predicate Abstraction}
\label{sec:predabs}

Predicate abstraction~\cite{ABE,AbstractionsFromProofs,HBMC-predicateabstraction}
abstracts the program's state space with predicates
that it learns using CEGAR~\cite{ClarkeCEGAR}
and Craig interpolation~\cite{AbstractionsFromProofs}.
Compared to symbolic execution, predicate abstraction is not limited to tracking
(symbolic) values and constraints in the program, but can derive more powerful abstractions.
The computation of abstractions can be costly,
thus predicate abstraction uses \emph{large-block encoding}~\cite{ABE,LBE} to
compute abstractions only at certain program locations,
which by default are the loop-head locations.
This reduces the number of abstractions calculated and, hence, the overall cost.
To run predicate abstraction, use the command \exonline{predicateAnalysis}:
\begin{Verbatim}[breaklines=true]
cpachecker --predicateAnalysis examples/example-safe.c
\end{Verbatim}
In this example, predicate abstraction derives the loop invariant \texttt{x + y == n},
which proves that \texttt{__assert_fail} in \cref{fig:ex-safe-c} is unreachable,
and hence returns the verdict \texttt{TRUE}.
Learned predicates at these locations are written down
in a format based on SMT-LIB2~\cite{SMTLIB2}
into the file \texttt{output/predmap.txt} of the current working directory.
Take the program in \cref{fig:ex-safe-c} for example.
Predicate abstraction can derive
the invariant \texttt{x + y == n} for the \texttt{while} loop at \cref{ex-safe-while}
in function~\texttt{main}
that suffices to prove the safety specification
that the assertion error is unreachable.
In \texttt{predmap.txt}, this is represented as follows:
\begin{Verbatim}[samepage=true]
(declare-fun |main::n| () (_ BitVec 32))
(declare-fun |main::y| () (_ BitVec 32))
(declare-fun |main::x| () (_ BitVec 32))

main:
(assert (= |main::n| (bvadd |main::y| |main::x|)))
\end{Verbatim}

Predicate abstraction can abstract the program state space
concisely in a way that proves the program safe, if it learns the right predicates.
Unfortunately,
there is no mechanism forcing predicate abstraction
to find predicates that abstract well.
Especially for concrete value assignments in the program,
the learned predicates might enumerate all possible states.
For instance, predicate abstraction may unnecessarily learn the predicates
\mbox{\texttt{x == 0}}, \mbox{\texttt{x == 1}}, and \mbox{\texttt{x == 2}}
at \cref{ex-const-while-a} of \cref{fig:ex-const},
instead of \texttt{z == 0}.
Alternatively, \impact~\cite{IMPACT} is another analysis that
abstracts a program's state space with predicates.
It computes and refines abstractions in a lazier way compared to predicate abstraction,
and can be initiated using the configuration \cpaconfig{predicateAnalysis-ImpactRefiner-ABEl}.
The two analyses have shown different and complementing strengths in our empirical evaluations~\cite{AlgorithmComparison-JAR}:
Predicate abstraction is more effective at deriving proofs,
whereas \impact is more efficient at finding specification violations.

\subsection{Bounded Model Checking}
\label{sec:bmc}

Bounded model checking (BMC)~\cite{BMC,AlgorithmComparison-JAR} is an analysis specialized in
finding specification violations.
Given a bound~$n$,
BMC symbolically unrolls the loops in the program $n$~times,
encodes all execution paths
and specification violations (within the unrolling bound~$n$) into an SMT formula,
and checks the satisfiability of the formula with an SMT~solver.
The satisfiability of the formula directly corresponds to the feasibility of the encoded error paths.
If the formula is satisfiable,
then a specification-violating execution path (with $n$~loop unrollings) exists and can be extracted from the satisfying assignment.
A bounded model checker then reports the verification verdict \texttt{FALSE}.
In case the formula is unsatisfiable,
the program is considered safe up to the bound~$n$.
A bounded model checker reports the verification verdict \texttt{TRUE}
if the loops in the program have finite bounds and are fully unrolled by the bound~$n$.
Otherwise, the verdict is \texttt{UNKNOWN},
as the behavior of the program at higher unrolling bounds is still unknown.

\cpachecker automatically determines the required unrolling bound by incrementally increasing the bound
using configuration \cpaconfig{bmc-incremental}.
Incremental BMC starts with an unrolling bound of~0 and
increments the bound by~1 after each iteration.
The analysis terminates once
an error path is found,
the safety specification is proven (by fully unrolling all loops in the program),
or a resource limit is reached.
For instance, the following command runs BMC with incrementally increasing loop bound
on the program in~\cref{fig:ex-unsafe-c} \exonline{bmc-unsafe}:
\begin{verbatim}
cpachecker --bmc-incremental examples/example-unsafe.c
\end{verbatim}
\cpachecker finds the bug inside the loop body of the program in~\cref{fig:ex-unsafe-c}
on its first encounter of the assertion, with zero complete unrollings of the loop.
Running incremental BMC on the correct program in~\cref{fig:ex-safe-c} does not succeed \exonline{bmc-safe}.
During the process, \cpachecker produces log messages that show the current unrolling bound:
\begin{Verbatim}[breaklines=True]
Adjusting maxLoopIterations to 2 (LoopBoundCPA:LoopBoundPrecisionAdjustment.nextState, INFO)
\end{Verbatim}
\cpachecker eventually reaches the time limit and the verdict is \texttt{UNKNOWN},
since a really large unrolling bound (roughly $2^{31}$) is required to fully explore the program.
If the loop condition at \cref{ex-safe-while} changes to \texttt{x > 0 \&\& x < 3} in~\cref{fig:ex-safe-c},
incremental BMC can prove the program safe with 2~complete loop unrollings.

\subsection{Extensions of BMC for Unbounded Verification}
\label{sec:bmc-extensions}

BMC can be extended for unbounded verification of programs
by employing the \kinduction principle~\cite{kInduction,InductionVerification}
or constructing fixed points, i.e., inductive invariants, via Craig interpolation~\cite{McMillanCraig,IMC-JAR,VizelFMCAD09,ForwardBackwardReachability}.
To run \kinduction in \cpachecker,
use the configuration \cpaconfig{kInduction},
which combines \kinduction with
an auxiliary invariant generator based on data-flow analysis~\cite{kInduction,CPA-DF} (described in~\cref{sec:df}).
The invariants produced by the latter are used to strengthen the induction hypotheses of the former.
This is more effective than plain \kinduction~\cite{kInduction}.
As opposed to incremental BMC,
\kinduction could easily prove the safety of the example programs in~\cref{fig:ex-safe-c}
with the command \exonline{kInduction}:
\begin{verbatim}
cpachecker --kInduction examples/example-safe.c
\end{verbatim}

\cpachecker has three verification algorithms based on BMC and Craig interpolation:
interpolation-based model checking (IMC)~\cite{McMillanCraig,IMC-JAR},
interpolation-sequence-based model checking (ISMC)~\cite{VizelFMCAD09,DAR-transferability},
and dual approximated reachability (DAR)~\cite{ForwardBackwardReachability,DAR-transferability}.
From unsatisfiable BMC queries,
the three algorithms derive interpolants to construct inductive invariants at loop heads.
Such an invariant overapproximates the reachable states of the program
that conforms to the safety specification,
and hence could serve as a proof for the program's correctness.
IMC, ISMC, and DAR are enabled via the configurations
\cpaconfig{bmc-interpolation},
\cpaconfig{bmc-interpolationSequence}, and
\cpaconfig{bmc-interpolationDualSequence}, respectively,
and currently support only programs with at most one loop.
The tool \cpachecker verifies the program in~\cref{fig:ex-safe-c} with IMC (\cpaconfig{bmc-interpolation})
via the command \exonline{bmc-interpolation}:
\begin{verbatim}
cpachecker --bmc-interpolation examples/example-safe.c
\end{verbatim}
It produces the below log message:
\begin{Verbatim}[breaklines=true]
The current image reaches a fixed point (IMCAlgorithm.reachFixedPointByInterpolation, INFO)
\end{Verbatim}
The message indicates that IMC has found an inductive invariant for the \texttt{while} loop at \cref{ex-safe-while}
and proved the safety specification of the program.

\subsection{Symbolic Memory Graphs with Symbolic Execution}
\label{sec:smg}

\cpachecker's symbolic-memory-graph (SMG) analysis~\cite{SMG}
combines symbolic execution~\cite{King76} with a graph-based domain that tracks all memory.
It is usable in \cpachecker with the configuration \cpaconfig{smg}.
In addition to common state-space exploration, the SMG analysis
can check for memory safety.
The analysis can detect memory leaks, invalid memory access, and invalid freeing of memory.

SMGs accurately track most memory operations, including
pointer arithmetics and bit-precise reading of memory.
They also store memory boundaries
and can thus be used to reason about the validity of pointer dereferences.
A distinguishing feature of SMGs
is that linked lists of arbitrary length can be abstracted under certain circumstances.
This is currently limited to lists that terminate in indefinitely repeating equal values.
If the analysis fails to abstract lists of arbitrary length,
it enumerates all possible list lengths.
This may lead to a path explosion, but can still find violations to safety specification.

\begin{figure}[t]
  \centering
\begin{lstlisting}[style=C]
#include <stdlib.h>
#include <(*@assert@*).h>
extern int __VERIFIER_nondet_int();
int main() {
  int size = 100;
  int num = __VERIFIER_nondet_int();
  int * arr = malloc(sizeof(int) * size);(*@\label{ex-memsafety-malloc}@*)
  for (int i = 0; i < size; i++) {
    arr[i] = num;(*@\label{ex-memsafety-assign}@*)
    num++;
  }
  for (int i = size; i >= 0; i--) {(*@\label{ex-memsafety-for2}@*)
    assert(*(arr + i) == num);(*@\label{ex-memsafety-assert}@*)
    num--;
  }
  return 0;(*@\label{ex-memsafety-return}@*)
}
\end{lstlisting}
  \vspace{-3mm}
  \caption{\exampleprog{example-unsafe-memsafety.c} with two distinct memory-safety violations}
  \label{fig:example-code-memsafety}
  \vspace{-5mm}
\end{figure}

We can see some capabilities of the SMG~analysis on the example program in \cref{fig:example-code-memsafety}.
The program first allocates some memory at \cref{ex-memsafety-malloc},
then uses this memory to store some distinct
but non-deterministic values in a loop at \cref{ex-memsafety-assign}, filling the entire memory allocated in \texttt{arr}.
Then, in a reversed loop,
the saved values are compared to their expected values at \cref{ex-memsafety-assert}.
Please note that this example is not pre-processed and
thus the command-line argument \texttt{-{}-preprocess} is needed.
To start the verification of memory safety with the configuration \cpaconfig{smg}
on this program, run the following command:
\begin{verbatim}
cpachecker --preprocess --smg --spec memorysafety \
    examples/example-unsafe-memsafety.c
\end{verbatim}
This detects that the first memory access of the second loop at \cref{ex-memsafety-for2} is unsafe (i.e., the verdict is \texttt{FALSE}),
as the pointer dereference exceeds the bounds of the allocated memory.
Another error can be found before \cref{ex-memsafety-return},
as the memory allocated in \texttt{arr} is never freed.
This second memory-safety violation can be found either
by fixing the invalid dereference at \cref{ex-memsafety-assert},
or by using the dedicated specification \cpaspec{memorycleanup}:
\begin{verbatim}
cpachecker --preprocess --smg --spec memorycleanup \
    examples/example-unsafe-memsafety.c
\end{verbatim}

\subsection{Termination Analysis}
\label{sec:termination}
\newcommand{\terminatingexampleprog}{\href{https://svn.sosy-lab.org/software/cpachecker/trunk/doc/examples/example-terminating.c?p=47297}{\texttt{example-terminating.c}}}
\newcommand{\nonterminatingexampleprog}{\href{https://svn.sosy-lab.org/software/cpachecker/trunk/doc/examples/example-nonterminating.c?p=47297}{\texttt{example-nonterminating.c}}}

The specification \emph{termination} requires a program to always terminate.
A program that can execute infinitely is called \emph{non-terminating}.

\cpachecker provides two approaches for termination analysis:
the termination-as-safety analysis~\cite{LivenessAsSafety} \cpaconfig{terminationToSafety}
and the lasso-based analysis~\cite{ULTIMATE-TERMINATION} \cpaconfig{lassoRankerAnalysis}.
Analysis \cpaconfig{terminationToSafety}
is based on loop unrolling (similar to BMC, cf.~\cref{sec:bmc}).
It can prove termination only if all loops in the program can be fully unrolled,
but is often efficient in finding specification violations,
i.e., counterexamples that show non-termination.
Analysis \cpaconfig{lassoRankerAnalysis}
constructs ranking functions and does not need to unroll
all loops in the program for termination proofs.

\newsavebox{\exampleTerminatingCode}
\begin{lrbox}{\exampleTerminatingCode}
  \begin{minipage}[b]{0.46\textwidth}
    \centering
\begin{lstlisting}[
      style=C,
      breaklines=true,
      %
    ]
extern unsigned __VERIFIER_nondet_uint();
int main() {
  unsigned int n = 1;
  unsigned int z = (*@\mbox{__VERIFIER_nondet_uint()}@*);
  while (n <= z) {(*@\label{ex-terminating-while}@*)
    (*@\textcolor{green}{n = n + 1;}@*)
    (*@\textcolor{green}{z = z - 1;}@*)
  }  
  return 0;
}
\end{lstlisting}
  \end{minipage}
\end{lrbox}

\newsavebox{\exampleNonterminatingCode}
\begin{lrbox}{\exampleNonterminatingCode}
  \begin{minipage}[b]{0.46\textwidth}
    \centering
\begin{lstlisting}[
      style=C,
      breaklines=true,
      %
    ]
extern unsigned __VERIFIER_nondet_uint();
int main() {
  int n = 1;
  int z = (*@\mbox{__VERIFIER_nondet_uint()}@*);(*@\label{ex-nonterminating-initz}@*)
  while (n <= z) {
    (*@\textcolor{red}{n = (n - 1) \% 3;}@*)
    (*@\textcolor{red}{z = (z + 1) \% 3;}@*)
  }  
  return 0;
}
\end{lstlisting}
  \end{minipage}
\end{lrbox}

\begin{figure}[t]
  \centering
  \subfloat[\terminatingexampleprog]{\usebox{\exampleTerminatingCode}\label{fig:ex-terminating-c}}
  \hfill
  \subfloat[\nonterminatingexampleprog]{\usebox{\exampleNonterminatingCode}\label{fig:ex-nonterminating-c}}
  \vspace{-3mm}
  \caption{Example C programs for demonstration of termination analyses}
  \label{fig:example-termination-code}
  \vspace{-5mm}
\end{figure}

\subsubsection{Termination-as-Safety Analysis}
The termination-as-safety analysis transforms
a verification task for a termination specification into
a verification task for reachability.
It stores the values of variables that were
seen at the programs' loop heads.
For example, the loop head for the two programs in \cref{fig:example-termination-code}
is the location that corresponds to~\cref{ex-terminating-while}.
Similar to BMC (cf. \cref{sec:bmc}), when the analysis visits a loop head
for the $n+1$-st time, it constructs an SMT formula that symbolically represents
$n$~loop unrollings.
Via satisfiability queries, the analysis checks whether there exists a reachable state
that is visited twice
within $n$~loop iterations.
If such a state is found, the program is non-terminating.

The following command line runs the analysis on the program
in \cref{fig:ex-nonterminating-c}
\exonline{terminationToSafety}:
\begin{verbatim}
cpachecker --terminationToSafety examples/example-nonterminating.c
\end{verbatim}
\cpachecker reports the verdict \texttt{FALSE}
and produces a counterexample
that shows the following three unrollings of the loop
(visible in the output file \texttt{output/\allowbreak Counterexample.1.core.txt}):
\begin{center}
  \texttt{(n, z)}: \texttt{(1, 2)} $\to$ \textcolor{red}{\texttt{\underline{(0, 0)}}} 
  $\to$ \texttt{(-1, 1)} $\to$ \texttt{(-2, 2)} $\to$ \textcolor{red}{\texttt{\underline{(0, 0)}}} 
\end{center}
The unrolling represents an execution with assignment \texttt{z\,=\,2} at \cref{ex-nonterminating-initz}.
By inspecting the values of \texttt{n} and \texttt{z}
at the loop-head location of each iteration,
we see that the state \texttt{(n,z)\,=\,(0,0)} is visited twice.
This represents a non-terminating loop.

\vspace{-0.2cm}
\subsubsection{Lasso-Based Analysis}
The main idea of the lasso-based analysis is to extract
potentially non-terminating structures called \emph{lassos} and then pass each of them to the
\href{https://www.ultimate-pa.org/?ui=tool\&tool=lasso\_ranker}{library \lassoranker}~\cite{RankingTemplates}.
This library constructs ranking functions, which are arguments for termination.
Simultaneously, it is looking for a non-termination argument. If it finds a non-termination argument for at least
one lasso, \cpachecker claims that the program is non-terminating.

The lasso-based analysis complements the termination-as-safety analysis.
The analysis can verify that program \terminatingexampleprog{} in \cref{fig:ex-terminating-c} terminates,
but not that program \nonterminatingexampleprog{} in \cref{fig:ex-nonterminating-c} might not terminate.
The following command line runs the lasso-based analysis on the program in \cref{fig:ex-terminating-c} \exonline{lassoRankerAnalysis}:
\begin{verbatim}
cpachecker --lassoRankerAnalysis examples/example-terminating.c
\end{verbatim}
\cpachecker reports the verdict \texttt{TRUE} and produces the
output file \texttt{output/termi\-nation\-Analysis\-Result.txt}.
This contains a termination argument in the form of the ranking function
$3*z - 3*n + 4$.
As $n$~is always positive,
if the loop condition $n \leq z$ is satisfied,
$3*z - 3*n + 4 \geq 0$ holds.
In addition, after each loop iteration,
the resulting value of the ranking function strictly decreases.
After a finite number of iterations,
the value will eventually become smaller than zero,
which implies the negation of the loop condition
and thus termination.

\vspace{-0.2cm}
\subsection{Integer-Overflow Detection}
\label{sec:overflow}

To detect integer overflows,
\cpachecker uses a standard reachability analysis, such as those explained in
\cref{sec:val,sec:predabs,sec:bmc},
together with an internal encoding of overflow conditions
as error locations (\cpachecker's overflow analysis also checks for underflows).
The configurations supporting overflow detection
have the suffix \verb|--overflow| in their names.
By default, \cpachecker only checks for signed integer overflows,
as these are declared undefined behavior by the C~standard.
To additionally check for unsigned integer overflows,
set the option \texttt{overflow.checkUnsigned} to \texttt{true}.
For instance, to determine whether the example program in \cref{fig:ex-safe-c}
is free of signed and unsigned integer overflows
while using predicate abstraction (cf. \cref{sec:predabs}),
run the command \exonline{predicateAnalysis-unsigned-overflow}:
\begin{Verbatim}[breaklines=true]
cpachecker --predicateAnalysis--overflow \
    --option overflow.checkUnsigned=true examples/example-safe.c
\end{Verbatim}
The verification verdict is \texttt{FALSE},
because an overflow could happen at \cref{ex-safe-inity} if
\texttt{n} and \texttt{x} are initialized to 0 and 1, respectively.

\vspace{-0.1cm}
\section{Conclusion}
\vspace{-0.1cm}

This tutorial gives an introduction to the \cpachecker framework
and how to use it to verify programs.
It gives an overview of the main analysis techniques that \cpachecker offers,
together with their strengths and weaknesses,
and provides guidance on how to use \cpachecker in several analysis situations.

We hope that our tutorial is useful for researchers, practitioners, and educators,
and that we stimulate interest and curiosity to dig deeper into the full potential
of software model checking.
Interested readers can find more information
on the \href{https://cpachecker.sosy-lab.org}{\cpachecker project web page},
in \href{https://cpachecker.sosy-lab.org/publications.php}{the research publications on \cpachecker},
the \href{https://gitlab.com/sosy-lab/software/cpachecker}{\cpachecker GitLab repository},
and the \href{https://groups.google.com/forum/#!forum/cpachecker-users}{\cpachecker mailing list}.

\inlineheadingbf{Data-Availability Statement}
\cpachecker is available at its project website
  {\small\url{https://cpachecker.sosy-lab.org}}
and Zenodo~\cite{CPAchecker-latest}.
This tutorial uses version~3.0~\cite{CPAchecker3.0}.
We also provide a reproduction package~\cite{CPAcheckerTutorial-artifact-FM24-proceedings}
that includes all the examples from this tutorial.

\inlineheadingbf{Funding Statement}
\cpachecker was funded in part
by the Canadian Natural Sciences and Engineering Research Council (NSERC)
--- \href{http://www.nserc-crsng.gc.ca/ase-oro/Details-Detailles_eng.asp?id=482301}{482301},
by the Deutsche Forschungsgemeinschaft (DFG)
--- \href{http://gepris.dfg.de/gepris/projekt/378803395}{378803395} (\href{https://convey.ifi.lmu.de/}{ConVeY}),
\href{http://gepris.dfg.de/gepris/projekt/418257054}{418257054} (\href{https://coop.sosy-lab.org/}{Coop}),
\href{http://gepris.dfg.de/gepris/projekt/496588242}{496588242} (\href{https://idefix.sosy-lab.org/}{IdeFix}),
\href{http://gepris.dfg.de/gepris/projekt/496852682}{496852682} (ReVeriX),
by the Free State of Bavaria, and by the LMU PostDoc Support Funds.

\newpage
\appendix

\section*{Appendices}
\section{List of Examples}
\label{sec:examples-list}

\Cref{tab:examples} repeats all examples from this tutorial
(directory prefix \texttt{examples/} for the example program is omitted for space reasons).
The example programs are provided
in our reproduction package~\cite{CPAcheckerTutorial-artifact-FM24-proceedings}
and
together with \cpachecker in the directory
\href{https://svn.sosy-lab.org/software/cpachecker/tags/cpachecker-3.0/doc/examples}{\texttt{doc/examples/}},
and are available online by clicking on the program name.
We also provide a \href{https://vcloud.sosy-lab.org/cpachecker/webclient/run/}{web interface for \cpachecker},
which has examples from the tutorial as selectable presets.

    {
        \small
        \setcounter{LTchunksize}{200}
        \rowcolors{0}{}{black!10}
        \begin{longtable}{p{3.3cm}<{\raggedright}@{\hspace{2.5mm}}p{6.7cm}<{\raggedright}@{\hspace{2.5mm}}l}
            \caption{\label{tab:examples}Overview of examples}\\[-1.5ex]
            \toprule
            \rowcolor{white}
            Example & Command line                                                                                                                                   & Ref. \\
            \midrule
            \endfirsthead
            \toprule
            \rowcolor{white}
            Example & Command line                                                                                                                                   & Ref. \\
            \midrule
            \endhead
            \bottomrule
            \multicolumn{3}{l}{{\scriptsize (continues on next page)}}\endfoot
            \bottomrule\endlastfoot
            \onWebclient{default}{Default analysis}
                    & \texttt{cpachecker \exampleprog{example-safe.c}}
                    & \cref{sec:hello-world}                                                                                                                                \\
            \onWebclient{assert}{Different specification}
                    & \texttt{cpachecker -{}-spec Assertion \exampleprog{example-safe.c}}
                    & \cref{sec:specification}                                                                                                                              \\
            \onWebclient{settingOptions}{Setting options}
                    & \texttt{cpachecker -{}-kInduction -{}-timelimit 900s
            -{}-heap 2000M -{}-spec ErrorLabel \exampleprog{example-safe.c}
            -{}-option solvers.solver=MATHSAT5}
                    & \cref{sec:config}                                                                                                                                     \\
            \onWebclient{valueAnalysis-NoCegar-join}{Value analysis without precision refinement and path-insensitive}
                    & \texttt{cpachecker -{}-valueAnalysis-NoCegar-join \exampleprog{example-const.c}}
                    & \cref{sec:val}                                                                                                                                        \\
            \onWebclient{dataFlowAnalysis}{Interval-based data-flow analysis}
                    & \texttt{cpachecker -{}-dataFlowAnalysis \exampleprog{example-const.c}}
                    & \cref{sec:df}                                                                                                                                         \\
            \onWebclient{symbolicExecution-NoCegar}{Symbolic execution with\-out precision refinement}
                    & \texttt{cpachecker -{}-symbolicExecution-NoCegar \exampleprog{example-sym.c}}
                    & \cref{sec:symex}                                                                                                                                      \\
            \onWebclient{symbolicExecution-Cegar}{Symbolic execution with precision refinement}
                    & \texttt{cpachecker -{}-symbolicExecution-Cegar \exampleprog{example-sym.c}}
                    & \cref{sec:symex}                                                                                                                                      \\
            \onWebclient{predicateAnalysis}{Predicate abstraction with \cegar}
                    & \texttt{cpachecker -{}-predicateAnalysis \exampleprog{example-safe.c}}
                    & \cref{sec:predabs}                                                                                                                                    \\
            \onWebclient{predicateAnalysis-ImpactRefiner-ABEl}{\impact}
                    & \texttt{cpachecker -{}-predicateAnalysis-ImpactRefiner-ABEl \exampleprog{example-unsafe.c}}
                    & \cref{sec:predabs}                                                                                                                                    \\
            \onWebclient{bmc-unsafe}{Incremental BMC on unsafe program}
                    & \texttt{cpachecker -{}-bmc-incremental \exampleprog{example-unsafe.c}}
                    & \cref{sec:bmc}                                                                                                                                        \\
            \onWebclient{bmc-safe}{Incremental BMC on safe program}
                    & \texttt{cpachecker -{}-bmc-incremental \exampleprog{example-safe.c}}
                    & \cref{sec:bmc}                                                                                                                                        \\
            \onWebclient{kInduction}{\kinduction with data-flow analysis}
                    & \texttt{cpachecker -{}-kInduction \exampleprog{example-safe.c}}
                    & \cref{sec:bmc-extensions}                                                                                                                             \\
            \onWebclient{bmc-interpolation}{Interpolation-based model checking}
                    & \texttt{cpachecker -{}-bmc-interpolation \exampleprog{example-safe.c}}
                    & \cref{sec:bmc-extensions}                                                                                                                             \\
            \onWebclient{bmc-interpolationSequence}{Interpolation-sequence-based model checking}
                    & \texttt{cpachecker -{}-bmc-interpolationSequence \exampleprog{example-safe.c}}
                    & \cref{sec:bmc-extensions}                                                                                                                             \\
            \onWebclient{bmc-interpolationDualSequence}{Dual approximated reachability}
                    & \texttt{cpachecker -{}-bmc-interpolationDualSequence \exampleprog{example-safe.c}}
                    & \cref{sec:bmc-extensions}                                                                                                                             \\
            Symbolic memory graphs with symbolic execution
                    & \texttt{cpachecker -{}-preprocess -{}-smg -{}-spec memorysafety \exampleprog{example-unsafe-memsafety.c}}
                    & \cref{sec:smg}                                                                                                                                        \\
            Symbolic memory graphs with symbolic execution for \cpaspec{memorycleanup}
                    & \texttt{cpachecker -{}-preprocess -{}-smg -{}-spec memorycleanup \exampleprog{example-unsafe-memsafety.c}}
                    & \cref{sec:smg}                                                                                                                                        \\
            \onWebclient{terminationToSafety}{Termination-as-safety analysis}
                    & \texttt{cpachecker -{}-terminationToSafety \nonterminatingexampleprog}
                    & \cref{sec:termination}                                                                                                                                \\
            \onWebclient{lassoRankerAnalysis}{Lasso-based termination analysis}
                    & \texttt{cpachecker -{}-lassoRankerAnalysis \terminatingexampleprog}
                    & \cref{sec:termination}                                                                                                                                \\
            \onWebclient{predicateAnalysis-overflow}{Overflow detection for signed integers}
                    & \texttt{cpachecker -{}-predicateAnalysis-{}-overflow \exampleprog{example-safe.c}}
                    & \cref{sec:overflow}                                                                                                                                   \\
            \onWebclient{predicateAnalysis-unsigned-overflow}{Overflow detection for signed and unsigned integers}
                    & \texttt{cpachecker -{}-predicateAnalysis-{}-overflow -{}-option overflow.checkUnsigned=true \exampleprog{example-safe.c}}
                    & \cref{sec:overflow}                                                                                                                                   \\
            Witness validation
                    & \texttt{cpachecker -{}-witnessValidation -{}-witness output/witness.yml \exampleprog{example-safe.c}}
                    & \cref{appendix:witness-validation}                                                                                                                    \\
            \onWebclient{testCaseGeneration-predicateAnalysis}{Test-case generation}
                    & \texttt{cpachecker -{}-testCaseGeneration-predicateAnalysis -{}-option testcase.xml=test-suite/testcase\%d.xml \exampleprog{example-unsafe.c}}
                    & \cref{appendix:test-gen}                                                                                                                              \\
        \end{longtable}
        \vspace{1em} %
    }

\section{Project Information}
\label{appendix:projectInfo}

\subsection{Development History}

The \cpachecker project was founded in
\href{https://svn.sosy-lab.org/trac/cpachecker/changeset/19155/CPAchecker}{2007}
to satisfy the demand for
a highly maintainable and extendable framework to explore various algorithms for software model checking and program analysis,
triggered on the one hand by the advent of CPAs, which suggests a modular architecture,
and on the other hand by the difficulty to maintain the \blast project~\cite{BLAST}
with its hard-coded design choices.
The initial milestone was to create an analyzer based on the parser from the
\href{https://projects.eclipse.org/projects/tools.cdt}{Eclipse CDT}
and a CPA for reaching definitions, which was implemented in a course project.
Originating from Simon Fraser University,
Universität Passau, and Ludwig-Maximilians-Universität München,
\cpachecker has received many contributions from other institutions,
such as Universität Paderborn,
Universität Oldenburg, Technische Universität Darmstadt,
Verimag in Grenoble, and ISP\,RAS.
The development history spans 17 years, and more than 120 developers worldwide
have contributed to its success.

\subsection{Achievements}
Over the past years,
submissions based on \cpachecker achieved top ranks in the international competitions
\svcomp~\cite{SVCOMP19,SVCOMP20,SVCOMP21,SVCOMP22,SVCOMP23,SVCOMP24},
\rers~\cite{RERS-STTT,RERS-History}, and
\testcomp~\cite{TESTCOMP19,TESTCOMP20,TESTCOMP21,TESTCOMP22,TESTCOMP23,TESTCOMP24}
(103~medals from SV-COMP, 13 medals from RERS, 4 medals from Test-Comp),
and \cpachecker received the Gödel medal in silver in 2014 for its contributions to the verification community
(see the \href{https://cpachecker.sosy-lab.org/achieve.php}{achievement page} of \cpachecker's website).
One \cpachecker paper~\cite{CPAexplicit} received a Test-of-Time Award from ETAPS.
The Linux Driver Verification project~\cite{LDV-Toolset,LDV,LDV12}
has successfully used \cpachecker to find
\href{https://svn.sosy-lab.org/software/cpachecker/tags/cpachecker-3.0/doc/Achievements.md#bugs-found-with-cpachecker}
{more than 240~bugs in Linux device drivers}.

\subsection{License and Reuse}

\cpachecker itself is open source and licensed under the \href{https://www.apache.org/licenses/LICENSE-2.0}{Apache 2.0 License}.
It includes the SMT solver \mathsat~\cite{MATHSAT5}
with a \href{https://svn.sosy-lab.org/software/cpachecker/tags/cpachecker-3.0/LICENSES/LicenseRef-MathSAT-CPAchecker.txt}{special license}
that allows use only for research and evaluation purposes.
Commercial use of \cpachecker can be done either with a different SMT solver
or an appropriate license from the \mathsat team.
(SMT solvers are integrated in \cpachecker via \javasmt~\cite{JavaSMT3};
 please refer to its GitHub page ({\smaller\url{https://github.com/sosy-lab/java-smt}}) for the available choices.)
All other bundled third-party components are available under standard open-source licenses.
Researchers as well as industrial users are welcome to use \cpachecker or integrate its components
into their workflow.

\section{Detailed Explanation of Verification Analyses}
\label{appendix:analysisDetails}

This section provides more detailed descriptions
of some verification analyses mentioned in \cref{sec:algorithms}
and how they work on the example programs.

\subsection{Example: Value Analysis}
\label{appendix:val}
\begin{figure}[t]
  \parbox{.46\linewidth}{%
    \hspace{0.075\linewidth}
\begin{lstlisting}[style=C]
extern void __assert_fail();
int main() {
  int x = 0;
  int y = 0;
  int z = 0;(*@\label{ex-const-initz}@*)
  while (x < 2) {(*@\label{ex-const-while}@*)
    x++;(*@\label{ex-const-xpp}@*)
    y = z + 1;(*@\label{ex-const-yz1}@*)
  }
  if (z != 0) {(*@\label{ex-const-if}@*)
    __assert_fail();
  }
  return 0;
}
\end{lstlisting}
  }\hfill%
  \parbox{.46\linewidth}{%
\centering
    \begin{tikzpicture}[
        argstate/.style={
            fill=gray!15!white, rounded corners=1pt, minimum width=10mm, minimum height=6mm,
            font=\scriptsize},
        argedge/.style={draw=black, ->},
        arglabel/.style={font=\small},
        node distance={3mm}
    ]

    \node[argstate] (1) {l.3: $x \mapsto 0$};
    \node[argstate, below=of 1] (2) {l.4: $x \mapsto 0, y \mapsto 0$};
    \node[argstate, below=of 2] (3) {l.5: $x \mapsto 0, y \mapsto 0, z \mapsto 0$};
    \node[argstate, below=of 3] (4) {l.6: $z \mapsto 0$};
    \node[argstate, below left=5mm and -5mm of 4] (loop1) {l.7: $z \mapsto 0$};
    \node[argstate, below=of loop1] (loop2) {l.8: $y \mapsto 1, z \mapsto 0$};
    \node[argstate, below right=5mm and -5mm of 4] (5) {l.10: $z \mapsto 0$};

    \path
        (1)     edge[argedge] (2)
        (2)     edge[argedge] (3)
        (3)     edge[argedge] (4)
        (4)     edge[argedge] (loop1)
        (4)     edge[argedge] (5)
        (loop1) edge[argedge] (loop2)
        (loop2.north west) edge[argedge,bend left=45] (4)
        ;
        
    \end{tikzpicture}
  }\\[3mm]
  \parbox[t]{.46\linewidth}{%
    \caption{Program~\exampleprog{example-const.c}
    (repeated from \cref{fig:ex-const} for easier reference)}
    \label{dup:ex-const}
  }\hfill%
  \parbox[t]{.46\linewidth}{%
    \caption{Abstract program states found by path-insensitive value analysis without precision refinement (\cpaconfig{valueAnalysis-NoCegar-join}) for the program in \cref{dup:ex-const}}
    \label{fig:ex-const-arg}
  }
\end{figure}%

\Cref{fig:ex-const-arg} shows the abstract program states
that the value-analysis configuration \cpaconfig{valueAnalysis-NoCegar-join}
(no path sensitivity, no precision refinement) computes
for the program in \cref{dup:ex-const},
to demonstrate how this technique trades precision for scalability.
Each state first lists the source-code line after that it holds, and then the abstract state
that is computed by this value-analysis configuration.
The states are in a predecessor-successor relationship:
A succeeding state is computed from one or more preceeding states.
For example abstract state l.5:~$x \mapsto 0, y \mapsto 0, z \mapsto 0$ says that
the value assignments $x \mapsto 0, y \mapsto 0, z \mapsto 0$ hold when the program execution
reaches line~5 from its predecessor state l.4:~$x \mapsto 0, y \mapsto 0$.
The value of $z$ is always~0, so it is always tracked.
The value of $y$ is always~0 at \cref{ex-const-initz} and always~1 at \cref{ex-const-yz1}, so it is kept for these locations.
But the value analysis does not track the value assignments
of $x$ beyond \cref{ex-const-initz}, because its value is not constant at \cref{ex-const-while,ex-const-xpp,ex-const-yz1,ex-const-if}:
It can be 0, 1, or~2.

With this approach, the value analysis can prove (on this example program)
properties over $z$ and partially over $y$, but not over $x$.

\subsection{Example: Check for Path Feasibility in Value Analysis}\label{appendix:value-analysis-cex-check}

\Cref{sec:val} explains the use of feasibility checks in the value analysis
of \cpachecker. These make sure that an error path
that is found by the rather imprecise value analysis
actually describes a specification violation.

During execution,
the following messages in the command-line output
show that \cpachecker{} performed a feasibility check that confirms a found error path:
\begin{Verbatim}[breaklines=True,fontsize=\small]
... snip other output ...
Error path found, starting counterexample check with CPACHECKER. (CounterexampleCheckAlgorithm.checkCounterexample, INFO)
... snip other output ...
Error path found and confirmed by counterexample check with CPACHECKER. (CounterexampleCheckAlgorithm.checkCounterexample, INFO)
\end{Verbatim}

And through the following messages it is visible that a check rejected the error path:
\begin{Verbatim}[breaklines=True,fontsize=\small]
Error path found but identified as infeasible. (CounterexampleCheckAlgorithm.checkCounterexample, INFO)

Infeasible counterexample found, but could not remove it from the ARG. Therefore, we cannot prove safety. (ExceptionHandlingAlgorithm.handleExceptionWithErrorPath, WARNING)

Stopping analysis ... (CPAchecker.runAlgorithm, INFO)

Analysis incomplete: no errors found, but not everything could be checked. (CPAchecker.analyzeResult, WARNING)

Verification result: UNKNOWN, incomplete analysis.
\end{Verbatim}
In the case of a rejected error path,
the value analysis continues to search for other (actual)
specification violations,
but it is not able to prove program safety anymore.
In these cases, the verification result can only be \texttt{FALSE} or \texttt{UNKNOWN}.

\subsection{Example: Symbolic Execution}
\label{appendix:symex}
\begin{figure}[t]
  \begin{minipage}[]{0.3\linewidth}
\begin{lstlisting}[breaklines=true,breakatwhitespace=true,style=C]
extern unsigned __VERIFIER_nondet_uint();
extern void __assert_fail();
int main() {
  unsigned int x = __VERIFIER_nondet_uint();
  unsigned int y = x;
  unsigned int z = __VERIFIER_nondet_uint();
  while (x < 2) {(*@\label{ex-symex-while}@*)
    x++;
    y++;
    z = x + z;
  }
  if (x != y) {(*@\label{ex-symex-if}@*)
    __assert_fail();
  }
  return 0;
}
\end{lstlisting}
    \caption{\raggedright Program \exampleprog{example-sym.c}
      (repeated from \cref{fig:ex-symex} for easier reference)}
    \label{dup:ex-symex}
    \vspace{2.2cm}
  \end{minipage}
  \hfill%
  \begin{minipage}[]{0.65\linewidth}
    \begin{tikzpicture}[
        argstate/.style={
            fill=gray!15!white, rounded corners=1pt, minimum width=10mm, minimum height=6mm,
            font=\scriptsize},
        argedge/.style={draw=black, ->},
        arglabel/.style={font=\small},
        node distance={3mm}
    ]

    \node[argstate] (0) {l.4: $\{x \mapsto s_1\}$, $\emptyset$};
    \node[argstate, below=of 0] (1) {l.5: $\{x \mapsto s_1, y \mapsto s_1\}$, $\emptyset$};
    \node[argstate, below=of 1] (2) {l.6: $\{x \mapsto s_1, y \mapsto s_1, z \mapsto s_2\}$, $\emptyset$};
    \node[argstate, below = of 2, anchor=north east, xshift=-2pt] (3) {l.7: $\{\ldots\}$, $\{s_1 < 2\}$};
    \node[argstate, below = of 2, anchor=north west, xshift=1pt] (4) {l.7: $\{\ldots\}$, $\{s_1 \geq 2\}$};
    \node[argstate, below = of 4, anchor=north west, xshift=1pt] (4-2) {l.12: \ldots};
    \node[argstate, below = of 3] (loop1) {l.8: $\{x \mapsto s_1+1, y \mapsto s_1, z \mapsto s_2\}$, $\{\ldots\}$};
    \node[argstate, below = of loop1] (loop1-2) {l.9: $\{x \mapsto s_1+1, y \mapsto s_1+1, z \mapsto s_2\}$, $\{\ldots\}$};
    \node[argstate, below = of loop1-2] (loop1-3) {l.10: $\{x \mapsto s_1+1, y \mapsto s_1+1, z \mapsto s_1 + 1 + s_2\}$, $\{\ldots\}$};
    \node[argstate, below=of loop1-3, anchor=north east, xshift=-3pt] (loop2) {l.7: $\{\ldots\}$, $\{s_1 < 2, s_1 + 1 < 2\}$};
    \node[argstate, below=of loop1-3, anchor=north west, xshift=-1pt] (5) {l.7: $\{\ldots\}$, $\{s_1 < 2, s_1 + 1 \geq 2\}$};
    \node[argstate, below = of 5] (5-2) {l.12: \ldots};

    \node[below = of loop2] (dots) {\vdots};

    \path
        (0)     edge[argedge] (1)
        (1)     edge[argedge] (2)
        (2)     edge[argedge] (3)
        (2)     edge[argedge] (4)
        (4)     edge[argedge] (4-2)
        (3)     edge[argedge] (loop1)
        (loop1) edge[argedge] (loop1-2)
        (loop1-2) edge[argedge] (loop1-3)
        (loop1-3) edge[argedge] (loop2)
        (loop1-3) edge[argedge] (5)
        (5) edge[argedge] (5-2)
        (loop2) edge[argedge] (dots)
        ;
        
    \end{tikzpicture}
    \caption{\raggedright Result of \cpachecker config \cpaconfig{symbolicExecution-NoCegar}
        on the program in \cref{dup:ex-symex}}
    \label{fig:ex-symex-arg}
  \end{minipage}
\end{figure}%

\Cref{fig:ex-symex-arg} shows an excerpt
of the symbolic-execution tree that symbolic execution
(cf.~\cref{sec:symex})
produces for the program in \cref{dup:ex-symex}.
For each non-deterministic choice
(modeled by a call to \texttt{__VERIFIER_nondet_uint()}), a new
symbolic value~$s_i$ is introduced.
When symbolic execution enters the \texttt{while} loop at \cref{ex-symex-while},
it tracks the constraint $s_1 < 2$ on symbolic value~$s_1$.
When it skips the loop and goes to \cref{ex-symex-if}, it tracks the constraint $s_1 \geq 2$.
Inside the loop it tracks, for each program operation, the expressions
over symbolic values: $s_1+1$ for variables~$x$ and $y$, and $s_1+1+s_2$ for~$z$.

Symbolic execution continues in this fashion until all program paths are fully explored.
\section{Verification-Witness Generation and Validation}\label{appendix:witness-validation}

A verification witness~\cite{WitnessesJournal,VerificationWitnesses-2.0} helps users reason about the verification result
and allows independent validation of the verification result.
\cpachecker can both export witnesses for verification results
and validate witnesses that other tools produce.

Witnesses encode information in a machine-readable format.
Currently, two formats exist for witnesses,
both of which are supported by \cpachecker:
version~1.0, the original GraphML-based format~\cite{WitnessesJournal},
and version~2.0, a newer YAML-based format~\cite{VerificationWitnesses-2.0}.
Witnesses version~2.0 are more succinct and human-readable than version~1.0,
but version~1.0 is, at the time of writing, still supported by more tools.
Due to this reason, \cpachecker can export witnesses in 
both formats.

Independent of the format,
witnesses are divided into two categories: \emph{correctness witnesses} for the verdict \texttt{TRUE}
and \emph{violation witnesses} for the verdict \texttt{FALSE}.

\subsection{Examples of Verification Witnesses}
\label{appendix:witnesses}

\newsavebox{\exampleUnsafeWitnessVAAppendix}
\begin{lrbox}{\exampleUnsafeWitnessVAAppendix}
        \begin{minipage}[b]{0.45\textwidth}
                \centering
                
\newcommand{\automatonfontsize}{\scriptsize}
\begin{tikzpicture}[
  ->,
  >=stealth',
  auto,
  node distance=0.4cm,
  semithick,
  font=\automatonfontsize,
  every initial by arrow/.style = {
    font=\normalfont\automatonfontsize,
  },
  initial distance=3mm,
  accepting/.style = {
    accepting by double,
  },
  every state/.style = {
    fill=white,
    draw=black,
    text=black,
    minimum size=7mm,
    inner sep=0pt,
    text height=1mm,
    text width=7mm,
    circle,
    align=center,
  },
  errorstate/.style = {
    state,
    accepting,
    fill=red!30,
    draw=red,
  },
  sinkstate/.style = {
    state,
    draw=blue,
    fill=blue!30,
  },
  every edge/.append style = {
    font=\ttfamily\automatonfontsize,
  },
 ]

  \node[white] at (-.5,0) {ble};
  \node[state] at ( 0.75, 0.00)    (0) [] {$\qinit$};
  \node[state]         at ( 0.75,-1.50)    (2) [] {$q_2$};
  \node[errorstate]    at ( 0.75,-3.00)   (E1) [] {$q_{E}$};

  \draw ($(0)+(0,0.6)$) to node {} (0);
  \path [every edge/.append style={text width=1cm}]
   (0) edge [pos=0.4,left, text width=2.5cm]                                        node[overlay] {\ref{ex-unsafe-inity}, x == 0 \&\& n == 1} (2)
       edge [loop right,text width=0.3cm,above,pos=.3]            node[xshift=-1.5mm] {\otherwise} (0)
   (2) edge [pos=0.4,left, text width=2.5cm]                                        node[overlay] {\ref{ex-unsafe-err}, true} (E1)
       edge [loop right,text width=0.3cm,above,pos=.3]            node[xshift=-1.5mm] {\otherwise} (2)
  (E1) edge [loop right,text width=0.3cm,above,pos=.3]             node[xshift=-1.5mm] {\otherwise} (E1);

\end{tikzpicture}

        \end{minipage}
\end{lrbox}

\newsavebox{\exampleSafeWitnessVAAppendix}
\begin{lrbox}{\exampleSafeWitnessVAAppendix}
        \begin{minipage}[b]{0.5\textwidth}
                \centering
                
\newcommand{\automatonfontsize}{\scriptsize}
\begin{tikzpicture}[
  ->,
  >=stealth',
  shorten >=1pt,
  auto,
  node distance=0.4cm,
  semithick,
  font=\automatonfontsize,
  every initial by arrow/.style = {
    font=\normalfont\automatonfontsize,
  },
  accepting/.style = {
    accepting by double,
  },
  every state/.style = {
    fill=white,
    draw=black,
    text=black,
    minimum size=7mm,
    inner sep=0pt,
    text height=1mm,
    text width=7mm,
    circle,
    align=center,
  },
  errorstate/.style = {
    state,
    accepting,
    fill=red!30,
    draw=red,
  },
  every edge/.append style = {
    font=\ttfamily\automatonfontsize,
  }
 ]

  \node[state,accepting,label=left:\text{\examplecodesize \formula{$\true$}{itpGreen}}]
    at (0.75, -1.75)   (1) [] {$q_1$};
  \node[state,accepting]
    at (0.75,-3.50)   (2) [] {$q_2$};
  \node[left=-0.3mm of 2,yshift=2.4mm] {\examplecodesize
    \formula{\hspace{-1mm}
      $x + y == n$\hspace{-1mm}
    }{itpGreen}};
  \node[state,accepting]
    at (-1.5,-5.25)   (3) [] {$q_3$};
    \node[left=0.3mm of 3] {\examplecodesize \formula{$\true$}{itpGreen}};
  \node[state,accepting]
    at (0.75,-5.25)   (4) [] {$q_4$};
    \node[right=0.3mm of 4] {\examplecodesize \formula{$\true$}{itpGreen}};

  \draw ($(1)+(0,0.6)$) to node {} (1);
  \path [every edge/.append style={text width=0.7cm}]

   (1) edge [pos=0.4,left,text width=2.2cm]      node {\ref{ex-safe-inity},enterLoopHead} (2)
       edge [loop right,text width=0.3cm,above]  node[yshift=1mm] {\hspace{-2mm}\otherwise} (1)
   (2) edge [loop right,text width=0.3cm,above]  node[yshift=1mm] {\hspace{-2mm}\otherwise} (2)
       edge [pos=0.4,left]                       node[xshift=-4mm] {\ref{ex-safe-while},then} (3)
       edge [pos=0.4,right,text width=0.7cm]     node {\ref{ex-safe-while},else} (4)
   (3) edge [loop right,text width=0.3cm,above]  node[yshift=1mm,xshift=-1.5mm] {\otherwise} (3)
       edge [out=320,in=235,pos=0.1,text width=0.3cm,below,looseness=1]
                                                 node[xshift=-5mm] {\ref{ex-safe-endwhile},enterLoopHead} (2)
   (4) edge [loop left,text width=0.3cm,above]   node[yshift=1mm,xshift=2mm] {\otherwise} (4);

\end{tikzpicture}

        \end{minipage}
\end{lrbox}

\begin{figure}[t]
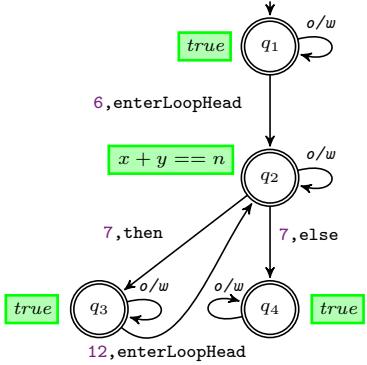
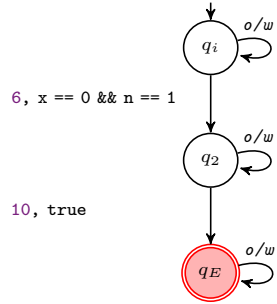

        \subfloat[Correctness witness for the the safe \\ program in~\cref{fig:ex-safe-c}]
        {\usebox{\exampleSafeWitnessVAAppendix}\label{fig:appendix:witnesses:correctness-witnessesv1}}
        \hfill
        \subfloat[Violation witness for the unsafe \\ program in~\cref{fig:ex-unsafe-c}]
        {\usebox{\exampleUnsafeWitnessVAAppendix}\label{fig:appendix:witnesses:violation-witnessesv1}}
        \caption{Conceptual view of example witnesses in version 1.0}
        \label{fig:appendix:witnesses:witnessesv1}
\end{figure}

\begin{figure}[t]
        \subfloat[Correctness witness for the safe program in~\cref{fig:ex-safe-c}]
        {
\begin{lstlisting}[language=XML,
                        morekeywords={node,edge,source,target,id},
                        numbers=none]
...
<node id="N1">
  <data key="entry">true</data>
</node>
<node id="N2"/>
<edge source="N1" target="N2">
  <data key="startline">2</data>
  <data key="endline">2</data>
  <data key="enterFunction"
    >main</data>
</edge>
<node id="N7">
  <data key="invariant">
      x + y == n</data>
  <data key="invariant.scope"
    >main</data>
</node>
<edge source="N2" target="N7">
  <data key="enterLoopHead"
    >true</data>
  <data key="startline">6</data>
  <data key="endline">6</data>
</edge>
...
\end{lstlisting}
                \label{fig:appendix:witnesses:correctness-witnessesv1:file}}
        \hfill
        \subfloat[Violation witness for the \\ unsafe program in~\cref{fig:ex-unsafe-c}]
        {
\begin{lstlisting}[language=XML,
                        morekeywords={node,edge,source,target,id},
                        numbers=none]
...
<node id="A0">
  <data key="entry">true</data>
</node>
<node id="A1"/>
<edge source="A0" target="A1">
  <data key="startline">2</data>
  <data key="endline">2</data>
  <data key="enterFunction"
    >main</data>
</edge>
<node id="A2"/>
<edge source="A1" target="A2">
  <data key="startline">6</data>
  <data key="endline">6</data>
  <data key="assumption"
    >x == 0 && n == 1</data>
  <data key="assumption.scope"
    >main</data>
</edge>



...
\end{lstlisting}
                \label{fig:appendix:witnesses:violation-witnessesv1:file}}
        \caption{Excerpt of the file contents for the example witnesses in version 1.0}
        \label{fig:appendix:witnesses:witnessesv1:file}
\end{figure}

\begin{figure}[t]
        \subfloat[Relevant sections of a correctness\\ witness for the safe program in~\cref{fig:ex-safe-c}]
        {\usebox{\exampleSafeWitnessVB}\label{fig:appendix:witnesses:correctness-witnessesv2}}
        \hfill
        \subfloat[Relevant sections of a violation \\ witness for the unsafe program in~\cref{fig:ex-unsafe-c}]
        {\usebox{\exampleUnsafeWitnessVB}\label{fig:appendix:witnesses:violation-witnessesv2}}
        \caption{Example witnesses in version 2.0
                (repeated from \cref{fig:witnesses:witnessesv2} for easier reference)}
        \label{fig:appendix:witnesses:witnessesv2}
\end{figure}

\Cref{fig:appendix:witnesses:witnessesv1} shows a conceptual view
and \cref{fig:appendix:witnesses:witnessesv1:file} the file contents
of examples of correctness and violation witnesses in version~1.0 for the
safe and unsafe C~programs in \cref{fig:example-code}, respectively,
and \cref{fig:appendix:witnesses:witnessesv2} does the same for version~2.0.
Similar witnesses can be generated with the following command:
\begin{verbatim}
cpachecker --predicateAnalysis examples/example-[un]safe.c 
  --option cpa.arg.proofWitness=witness.graphml
\end{verbatim}

In the case of correctness witnesses, version 1.0 contains information about
the control flow of the program. This information is not contained in the witness in version 2.0.
Still, both present the same fundamental information,
specifically, that the program
has a loop invariant \texttt{x + y == n} at the loop head in \cref{ex-safe-while}.
The invariant is inductive, and therefore it is straightforward to
proof the program correct using it.

In the case of violation witnesses, both witnesses describe a set of paths through the program
that lead to the violation of the specification. This makes it easier
to replay the violation, since only the paths restricted by
the witness need to be considered. For the presented
witnesses the set of paths consists of a single one, which
is given by the path which occurs when
\texttt{n == 1} and \texttt{x == 0} are assumed before \cref{ex-unsafe-inity}.
The witnesses state that if this path is followed then there will be
a violation of the specification at \cref{ex-unsafe-err}.

\subsection{Verification-Witness Generation}
By default, \cpachecker exports violation witnesses in both version 1.0 
and 2.0 and correctness witnesses in version 2.0.
If output files are disabled explicitly,
e.g., by the command-line argument \texttt{-{}-benchmark},
it is possible to specifically enable witness exporting
with the following options.
To export a verification witness in version~1.0
to the file \texttt{output/witness.graphml}
of the current working directory,
the three options
\texttt{cpa.arg.proofWitness},
\texttt{counterexample.export.graphml}, and
\texttt{termination.violation.witness}
must all be set to \texttt{witness.graphml}.
The last option configures the output files for non-termination witnesses,
whereas the first and second configure the output for
correctness and violation witnesses for other specifications, respectively.
By default, the exported witnesses in version 1.0 
for specifications other than termination
are compressed by \gzip (with file extension \texttt{.gz}).
To disable this, set the options \texttt{cpa.arg.compressWitness}
and \texttt{counterexample.export.compressWitness} to \texttt{false}.
Similarly, to export a witness in version~2.0 to the
file \texttt{output/witness.yml}
of the current working directory,
the two options
\texttt{cpa.arg.yamlProofWitness} and
\texttt{counterexample.export.yaml}
should both be set to \texttt{witness.yml}.
To make \cpachecker export witnesses in both versions,
provide all options simultaneously.

\subsection{Verification-Witness Validation}
\cpachecker's witness validation~\cite{WitnessesJournal}
tries to recompute a claimed verification verdict
with the help of the information that is provided by a verification witness.
The following command line validates the verdict of \cpachecker's successful verification of program~\exampleprog{example-safe.c}~(\cref{fig:ex-safe-c})
using the witness of \cref{fig:appendix:witnesses:correctness-witnessesv2} (here stored as file \texttt{witness.yml}).
\begin{verbatim}
cpachecker --witnessValidation --witness witness.yml \
  examples/example-safe.c
\end{verbatim}
The configuration \cpaconfig{witnessValidation} enables witness validation and the
command-line argument \texttt{-{}-witness witness.yml} provides the verification witness.
Last, \texttt{example-safe.c} defines the program under analysis for that the witness claims the
corresponding verification result.

For the above command line, \cpachecker reports the verdict~\texttt{TRUE}.
This verdict must be interpreted based on the witness type:
If \cpachecker produces the same verdict that the verification witness claims
(\texttt{TRUE} for correctness witnesses and \texttt{FALSE} for violation witnesses),
it \emph{confirms} the witness.
In the example, the provided witness is a correctness witness and \cpachecker reports \texttt{TRUE}.
This means that the witness is confirmed.

By default, \cpachecker checks that a witness's metadata is consistent with the program-under-analysis.
In case you want to validate a witness whose metadata does not match
the program-under-analysis,
use \texttt{witness.strictChecking=false} which disables metadata checks.

Next, we discuss how \cpachecker validates witnesses,
first for violation witnesses.
Except for witnesses for non-termination,
we need to confirm a violation of a safety specification that can be explained with a finite execution path.
Therefore, we expect the witness to describe at least one finite execution path.
During validation, we use the provided witness to restrict the state-space exploration to the paths
described by the witness.
To enforce the restrictions, we first convert the witness into a control automaton,
which is similar to the specification automata described in~\cref{sec:specification}
but allows restricting the analyzed state space.
Thereafter, we run an appropriate analysis in composition with the control automaton to explore the relevant paths.
\cpachecker confirms a specification violation if the analysis reaches a violation state that is also marked as a violation in the witness.
For sequential programs, we mainly rely on predicate abstraction~\cite{ABE,AlgorithmComparison-JAR} to check the (un)reachability of certain error locations or no-overflow specifications (cf. \cref{sec:predabs}).
To support recursion, we enable block abstraction memoization~\cite{BAM}.
To check memory-related specifications, we apply our SMG analysis~\cite{SMG} (cf. \cref{sec:smg}).
For concurrent programs, we use a particular BDD-based analysis~\cite{CPABDD-memics,ThreadingCPA}.
Witnesses that describe non-termination cannot be explained with a finite execution path,
but only with an infinite one.
Our validation expects that the infinite execution is described by a path to a loop and the loop itself, which is characterized by a quasi loop invariant (i.e., it ensures that when it becomes true it remains true).
Thus, two conditions need to be checked:
(1)~the path to the loop indeed reaches the loop in a state fulfilling the quasi loop invariant,
and (2)~if we start a loop iteration in a state fulfilling the quasi loop invariant, then we do not leave the loop and at the end of the iteration we are in a state that also fulfills the quasi loop invariant.
During validation, we express both checks as reachability analyses performed with the help of predicate abstraction~\cite{ABE,AlgorithmComparison-JAR}.

For validation of correctness witnesses, \cpachecker expects the witness to provide loop invariants.
\cpachecker checks their validity during analysis.
In case of (un)reachability specifications,
we extract the invariants from the witness and perform \kinduction~\cite{kInduction} with candidate invariants from the witness (cf. \cref{sec:bmc-extensions}).
If an invariant does not hold for the current~$k$ in \kinduction, it is ignored in that step
and retried for $k+1$. %
In this mode, \cpachecker does not add any other candidate invariants.
This means that the correctness proof fully relies on the loop invariants provided by the witness
to be $k$-inductive for some (ideally small)~$k$.
For the no-overflow specification, we use predicate abstraction~\cite{ABE,AlgorithmComparison-JAR} and enable block-abstraction memoization~\cite{BAM} for recursive programs.
In parallel to the predicate abstraction, we check whether the invariants provided in the witness are valid.
If an invariant is proved invalid the validation reaches a violation, otherwise predicate abstraction assumes the invariant.
In the end, \cpachecker confirms the correctness witness
if the analyses encounter no specification violation.

\section{Test-Suite Generation}
\label{appendix:test-gen}

As mentioned in~\cref{sec:output-tests},
\cpachecker typically outputs a matching test harness
if it finds a specification violation.
However, \cpachecker can also
generate whole test suites
that cover paths to specific target locations
or fulfill a structural coverage specification
such as branch coverage.

\subsection{Running Test-Suite Generation}

The following command line generates a test suite
for the example program in \cref{fig:ex-unsafe-c}
which fulfills the default coverage specification (branch coverage). 
It uses \cpachecker's predicate analysis (cf.~\cref{sec:predabs}) for this
\exonline{testCaseGeneration-predicateAnalysis}.
\begin{verbatim}
cpachecker --testCaseGeneration-predicateAnalysis \
  --option testcase.xml=test-suite/testcase%d.xml \
  examples/example-unsafe.c
\end{verbatim}
Configuration option \texttt{testcase.xml=test-suite/testcase\%d.xml} defines
that the generated test cases are written in the exchange format of Test-Comp as files \texttt{testcase\{0,1,2,3\}.xml}
in the directory \texttt{output/test-suite/}.

The program in \cref{fig:ex-unsafe-c} has two branching locations, namely the conditions of the while loop at \cref{ex-unsafe-while} and the if statement at \cref{ex-unsafe-if},
which results in four test goals for branch coverage.
On the console, \cpachecker prints the following log message,
which shows that the analysis has achieved \SI{100}{\%} branch coverage
(log level \texttt{SEVERE} is expected for this output):
\begin{verbatim}
4 of 4 covered (TestCaseGeneratorAlgorithm.run, SEVERE)
\end{verbatim}
Furthermore, \cpachecker reports \texttt{Finished}.

As an example, the test case \texttt{output/test-suite/testcase0.xml}
that is produced by the above command line has the following content:
\begin{verbatim}
<testcase>
  <input>2U</input>
  <input>2U</input>
</testcase>
        \end{verbatim}

\begin{table}[t]
    \caption{Supported output formats for test cases}
    \label{tab:test-case-formats}
    \centering
    \rowcolors{0}{}{black!10}
    \begin{tabular}{p{0.43\textwidth} p{0.56\textwidth}}
        \toprule
        Format                                            & Configuration option with recommended value      \\
        \midrule
        Test vector as plain text                         & \verb|testcase.values=test-suite/testcase%d.txt| \\
        Test vector in \testcomp~\cite{TESTCOMP23} format & \verb|testcase.xml=test-suite/testcase%d.xml|    \\
        Test harness as compilable C code                 & \verb|testcase.file=test-suite/testcase%d.c|     \\
        \bottomrule
    \end{tabular}
\end{table}

\subsection{Configuring Test-Case Export}
Depending on the configuration, the exported test case is output in one or more of the formats
listed in~\cref{tab:test-case-formats}.
By default the test cases contain input values for all external function calls
except for standard functions like \texttt{malloc}, \texttt{exit}, etc.
and for both external and used, but uninitialized variables.
The option \texttt{testcase.excludeInitialization=true} makes \cpachecker
exclude the values for external and uninitialized variables from the test cases,
and the option \texttt{testHarnessExport.onlyVerifierNondet=true}
makes \cpachecker exclude all input values but those for
calls to external functions that are named \mbox{\texttt{\_\_VERIFIER\_nondet*}}.
\cpachecker may fail to compute sane input values for pointers, composite structures, or array initialization.
In such cases, it uses dummy values.

\begin{table}[t]
    \vspace{-0.5cm} %
    \caption{Supported structural coverage properties}
    \label{tab:coverageProperties}
    \centering
    \rowcolors{0}{}{black!10}
    \begin{tabular}{p{0.35\textwidth} p{0.64\textwidth}}
        \toprule
        Coverage criterion                   & Configuration option                                                \\
        \midrule
        Call to a specified function f       & \texttt{testcase.targets.type=FUN\_CALL testcase.targets.funName=f} \\
        Call to \texttt{\_\_VERIFIER\_error} & \texttt{testcase.targets.type=ERROR\_CALL}                          \\ %
        Statement coverage                   & \texttt{testcase.targets.type=STATEMENT}                            \\
        Branch coverage                      & \texttt{testcase.targets.type=ASSUME}                               \\
        \bottomrule
    \end{tabular}
\end{table}

\subsection{Configuring Test-Suite Generation}
The \texttt{testCaseGeneration-*} that are
provided in \cpachecker's \href{https://svn.sosy-lab.org/software/cpachecker/tags/cpachecker-3.0/config}{\texttt{config/}} directory
make \cpachecker generate test cases for structural properties like branch coverage.
Note that some configurations combine different analyses, as for example done by \coveritest~\cite{COVERITEST,CoVeriTest-STTT}.

It is sufficient to call \cpachecker with the selected configuration and specify
the desired test-case format (explained in the previous section) and
the coverage property.
\Cref{tab:coverageProperties} gives an overview of the supported coverage properties and their specification. %
For example,
the following command
generates a test case that reaches the function \texttt{__assert_fail} in the program in \cref{fig:ex-unsafe-c}
\exonline{testCaseGeneration-settings}:
\begin{verbatim}
cpachecker --testCaseGeneration-predicateAnalysis \
  --option testcase.targets.type=FUN_CALL \
  --option testcase.targets.funName=__assert_fail \
  --option testcase.xml=test-suite/testcase%d.xml \
  examples/example-unsafe.c
\end{verbatim}
\cpachecker will print on the console
that the test goal is covered by the analysis,
and output the corresponding test case to the output directory.

\bibliography{bib/dbeyer,bib/sw,bib/artifacts,bib/svcomp,bib/svcomp-artifacts,bib/testcomp,bib/testcomp-artifacts,bib/websites}

\end{document}